\documentclass[acmsmall]{acmart}
\usepackage{xcolor}
\usepackage{enumitem}
\usepackage{booktabs}
\usepackage{arydshln}
\usepackage{amsmath}
\usepackage{bbm}
\usepackage{bm}
\usepackage[ruled,vlined]{algorithm2e}
\usepackage{url}
\usepackage{float}
\usepackage[export]{adjustbox}
\usepackage{marvosym}
\usepackage{pifont}
\usepackage{multirow}
\usepackage{mathtools}
\usepackage{cancel}
\usepackage{stfloats}
\usepackage{multirow}
\usepackage{color, colortbl}

\DeclareMathOperator*{\argmax}{arg\,max}

\definecolor{Gray}{gray}{0.9}
\definecolor{DarkGreen}{rgb}{0.01, 0.75, 0.24}

\AtBeginDocument{%
  \providecommand\BibTeX{{%
    \normalfont B\kern-0.5em{\scshape i\kern-0.25em b}\kern-0.8em\TeX}}}

\acmDOI{XXXXXXX.XXXXXXX}
\acmJournal{TIST}

\begin{document}

\title{ReuseKNN: Neighborhood Reuse for Differentially-Private KNN-Based Recommendations}

\author{Peter Müllner}
\affiliation{%
  \institution{Know-Center GmbH and Graz University of Technology}
  \city{Graz}
  \country{Austria}}
\email{pmuellner@know-center.at}
\email{pmuellner@student.tugraz.at}

\author{Elisabeth Lex}
\affiliation{%
  \institution{Graz University of Technology}
  \city{Graz}
  \country{Austria}}
 \email{elisabeth.lex@tugraz.at}
 
\author{Markus Schedl}
\affiliation{%
  \institution{Johannes Kepler University Linz and Linz Institute of Technology}
  \city{Linz}
  \country{Austria}}
 \email{markus.schedl@jku.at}

\author{Dominik Kowald}
\affiliation{%
  \institution{Know-Center GmbH and Graz University of Technology}
  \city{Graz}
  \country{Austria}}
\email{dkowald@know-center.at}
\email{dominik.kowald@tugraz.at}

\renewcommand{\shortauthors}{Müllner P., Lex, E., Schedl M., and Kowald, D.}

\begin{abstract}
  User-based \emph{KNN} recommender systems (\emph{UserKNN}) utilize the rating data of a target user's $k$ nearest neighbors in the recommendation process.
    This, however, increases the privacy risk of the neighbors, since the recommendations could expose the neighbors' rating data to other users or malicious parties.
    To reduce this risk, existing work applies differential privacy by adding randomness to the neighbors' ratings, which unfortunately reduces the accuracy of \emph{UserKNN}.
    In this work, we introduce \emph{ReuseKNN}, a novel differentially-private KNN-based recommender system. 
    The main idea is to identify small but highly reusable neighborhoods so that (i) only a minimal set of users requires protection with differential privacy, and (ii) most users do not need to be protected with differential privacy, since they are only rarely exploited as neighbors. 
    In our experiments on five diverse datasets, 
    we make two key observations: Firstly, \emph{ReuseKNN} requires significantly smaller neighborhoods, and thus, fewer neighbors need to be protected with differential privacy compared to traditional \emph{UserKNN}.
    Secondly, despite the small neighborhoods, \emph{ReuseKNN} outperforms \emph{UserKNN} and a fully differentially private approach in terms of accuracy.
    Overall, \emph{ReuseKNN} leads to significantly less privacy risk for users than in the case of \emph{UserKNN}.
\end{abstract}



\maketitle

\section{Introduction}
\begin{figure}[!t]
    \centering
    \includegraphics[width=1\linewidth]{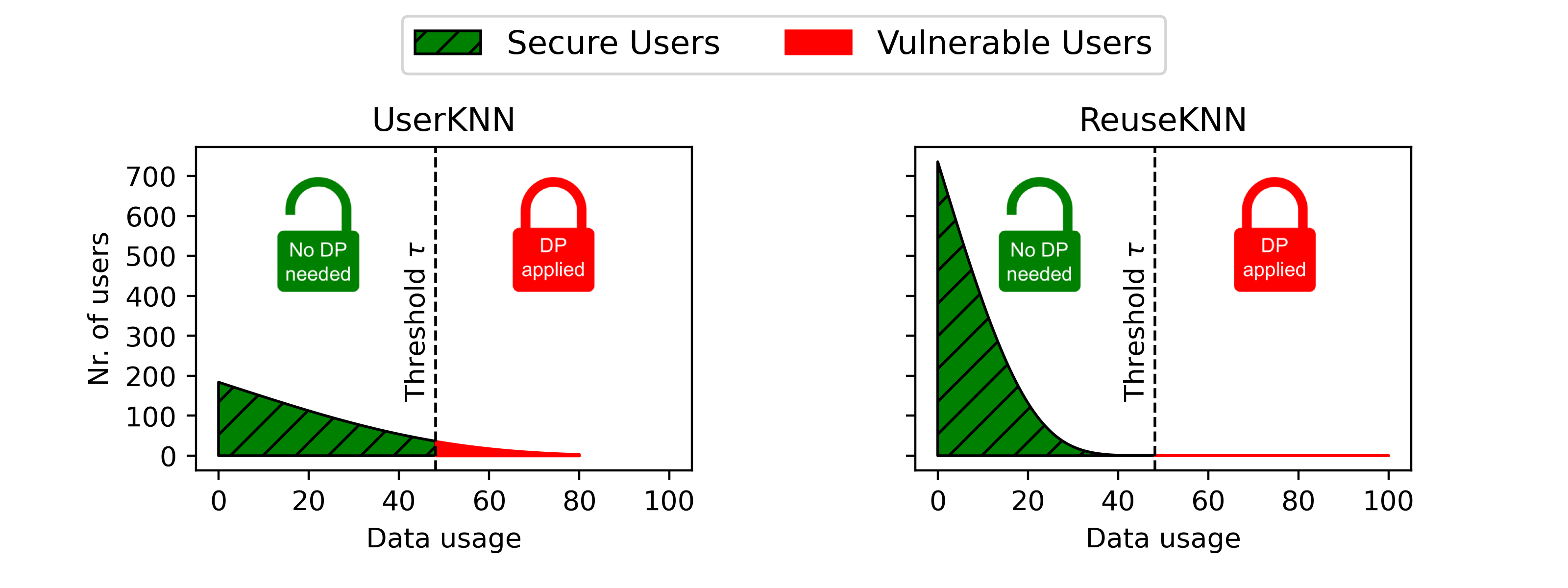}
    \caption{Schematic illustration of the data usage (i.e., how often a user is used as a neighbor) distribution of traditional \emph{UserKNN} and the proposed \emph{ReuseKNN} recommender system. 
    \emph{ReuseKNN} increases the number of secure users (green, no differential privacy needed) and decreases the number of vulnerable users (red, differential privacy needs to be applied) compared to \emph{UserKNN}.
    The dashed line illustrates the data usage threshold $\tau$, a hyperparameter for adjusting the maximum data usage for a user to be treated as secure.}
    \label{fig:approach}
\end{figure}
 Recommender systems often rely on {neighborhood-based} collaborative filtering~\cite{herlocker1999algorithmic} to generate recommendations.
{Such recommender systems can intuitively justify their recommendations to the target user and also, efficiently incorporate new rating data from users, which are two key issues of modern recommender systems~\cite{desrosiers2010comprehensive}.}
{For example,} user-based \emph{KNN}, i.e., \emph{UserKNN}, is a variant of {neighborhood-based} collaborative filtering that utilizes the rating data of the $k$ nearest neighbors of a target user to process a rating query. 
A rating query is a request to a recommender system to predict a rating for a target user to a target item.
However, the way how rating queries are processed by \emph{UserKNN} can increase the privacy risk of users since the estimated rating scores, which determine whether an item will be recommended, are generated based on rating data of users that are used as neighbors. 
In this regard, existing research~\cite{ramakrishnan2001being,calandrino2011you,zhang2021membership} finds that these neighbors are susceptible to multiple privacy risks, such as the inference of their private rating data (see Section~\ref{sec:problem}). 
To mitigate that privacy risk, several works~\cite{10.1145/3501812,gao2020dplcf,zhu2013differential} use \emph{differential privacy (DP)}~\cite{dwork2014algorithmic, dwork2008differential} to protect users' rating data by adding a degree of randomness to the data.
However, the added randomness typically leads to severe drops in recommendation accuracy~\cite{berkovsky2012impact}.

To address this problem, we introduce \emph{ReuseKNN}, a novel differentially-private KNN-based recommender system that reduces the number of neighbors to which differential privacy needs to be applied. 
Intuitively, instead of utilizing new users as neighbors for processing new rating queries, \emph{ReuseKNN} reuses useful neighbors from past rating queries.
Hence, \emph{ReuseKNN} constructs small but highly reusable neighborhoods for every target user by fostering the neighbors' reusability for many rating queries.
With this, as illustrated in  Figure~\ref{fig:approach}, \emph{ReuseKNN} minimizes the set of users that need to be protected with DP - we call them ``vulnerable users''.
Plus, most users do not need to be protected with DP, as their rating data is only rarely used in the recommendation process -  we call them ``secure users''.
As shown, we also introduce a data usage threshold $\tau$, i.e., a hyperparameter that allows adjusting the maximum data usage for a user to be treated as secure.
In this way, we leave it to the recommender system provider to specify what degree of data usage is tolerated despite the resulting privacy risks, and which users need to be protected. 

We evaluate the proposed approach in a two-stage procedure: (i) neighborhood reuse only, i.e., \emph{ReuseKNN}, and (ii) neighborhood reuse with DP, i.e., \emph{ReuseKNN}$_{DP}$.
In the first stage, \emph{ReuseKNN} does not use DP at all.
With this, we focus on how neighborhood reuse can increase the reusability of neighbors and preserve \emph{UserKNN}'s recommendation accuracy.
In the second stage, we combine \emph{ReuseKNN} with DP, i.e., \emph{ReuseKNN}$_{DP}$, to protect vulnerable users with DP.
This allows investigating how \emph{ReuseKNN}$_{DP}$ can mitigate all users' privacy risk while generating accurate recommendations.
We evaluate \emph{ReuseKNN} and \emph{ReuseKNN}$_{DP}$ on five different datasets, i.e., \emph{MovieLens 1M}, \emph{Douban}, \emph{LastFM}, \emph{Ciao}, and \emph{Goodreads}.
Plus, we compare \emph{ReuseKNN} and \emph{ReuseKNN}$_{DP}$ to five KNN-based baselines that utilize DP (e.g.,~\cite{zhu2013differential}) and the concept of neighborhood reuse in different ways, with respect to recommendation accuracy and users' privacy risk.
Additionally, the nature of neighborhood reuse may raise concerns that the generated recommendations are biased towards items consumed by many users, i.e., popular items.  
Thus, we investigate whether the proposed approach is more or less prone to item popularity bias than the baselines.

Our results indicate that \emph{ReuseKNN} yields significantly smaller neighborhoods than traditional \emph{UserKNN}.
Despite the smaller neighborhoods, \emph{ReuseKNN} and \emph{ReuseKNN}$_{DP}$ outperform our baselines in terms of recommendation accuracy.
Moreover, \emph{ReuseKNN}$_{DP}$ leads to significantly less privacy risk for users than \emph{UserKNN} with DP.
Also, the proposed approach does not increase item popularity bias.
Overall, the three main contributions of this paper are as follows:
\begin{enumerate}
    \item
    We present a novel \emph{ReuseKNN} recommender system and compare two neighborhood reuse strategies to substantially foster the reusability of a target user's neighborhood and effectively reduce the number of vulnerable users.
    \item
    We combine \emph{ReuseKNN} with DP to realize \emph{ReuseKNN}$_{DP}$ and show that \emph{ReuseKNN}$_{DP}$ improves recommendation accuracy over KNN- and DP-based baselines, and at the same time, does not increase item popularity bias.
    \item
    We show that \emph{ReuseKNN}$_{DP}$ leads to significantly less privacy risk, since most users are rarely exploited in the recommendation process and only the remaining users, i.e., vulnerable users, are protected with DP.
\end{enumerate}
Our work illustrates how to address privacy risks in KNN-based recommender systems through neighborhood reuse combined with DP.
While the proposed approach focuses on traditional \emph{KNN}, we additionally demonstrate the generalizability of the neighborhood reuse principle to user and item embeddings created by state-of-the-art neural collaborative filtering approaches~\cite{he2017neural}. 

\section{Related Work}
\label{sec:related_work_and_background}
We describe two research strands related to our work: (i) studies on the identification and quantification of users' privacy risks in recommender systems, and (ii) privacy-aware recommender systems that mitigate users' privacy risks. 
Since \emph{ReuseKNN} is a differentially-private and KNN-based recommender system, we emphasize KNN-based methods when reviewing privacy risks in recommender systems as well as DP when reviewing privacy-preserving technologies for recommender systems.
Also, we focus on the privacy risks that arise from the recommendations presented to potentially malicious target users.
This can harm the neighbors used in the recommendation process. 

\subsection{Privacy Risks in Recommender Systems} 
Previous research~\cite{beigi2020survey,jeckmans2013privacy,friedman2015privacy, ramakrishnan2001being} 
describes many severe privacy risks for users of recommender systems. 
For example, according to Ramakrishnan et al.~\cite{ramakrishnan2001being}, the use of neighbors' rating data in the recommendation process can pose a privacy risk to the neighbors.
Serendipitous recommendations could reveal unique connections between neighbors and items.
In this way, the rating data of the neighbors can be uncovered, or the neighbors' identities can be revealed within the recommendation database.
Also, Zhang et al.~\cite{zhang2021membership} show that it could be possible to identify users whose data was used in the recommendation process. 
Their results suggest that the effectiveness of their attack depends on the number of generated recommendations. 
Moreover, Calandrino et al.~\cite{calandrino2011you} propose to generate fake users, i.e., sybils, based on limited knowledge of a victim's data.
These sybils can isolate the victim that is utilized as a neighbor and compromise its privacy.


To quantify users' privacy risks in computational systems such as recommender systems, several privacy risk metrics~\cite{chen2020fine,domingo2010rational,wagner2018technical,liu2010framework,srivastava2013measuring} have been proposed.
These metrics often rely on the sensitivity of users' data, i.e., how strong this data puts users' privacy at risk.
For example, Chen et al.~\cite{chen2020fine} detect correlations within the dataset to measure if a piece of data could reveal personal information about the users.
Srivastava et al.~\cite{srivastava2013measuring} measure the relative sensitivity of a single piece of data compared to the remaining data of a user.
Similarly, Domingo-Ferrer~\cite{domingo2010rational} relates the overall sensitivity of a user's data to the sensitivity of other users' data. 
Besides, Liu and Terzi's \emph{privacy score}~\cite{liu2010framework} weighs the sensitivity with the degree of visibility of a user's data (i.e., how often a user's data is utilized in the recommendation process).

Evaluating the privacy risk of users based on attacks only measures the privacy risk with respect to the specific attack scenario.
Liu and Terzi's metric measures users' privacy risk independent of specific attack scenarios and, thus, allows investigating privacy risk in a recommender system at a more general level.
Therefore, in our work, we utilize Liu and Terzi's metric to measure users' privacy risk in a general neighborhood-based recommendation scenario. 
Furthermore, we assume that all pieces of data are equally sensitive, since sensitivity typically depends on the application and the user's perception of privacy~\cite{knijnenburg2013making}.

\subsection{Privacy-Aware Recommender Systems} 
Several works~\cite{10.1145/2508037.2508048,zhang2021privacy,tang2016privacy} mitigate users' privacy risks by applying \emph{homomorphic encryption}~\cite{gentry2009fully} to users' rating data.
Here, recommendations are generated based on the encrypted rating data, and thus, users' rating data remains protected in the recommendation process.
Homomorphic encryption, however, has high computational complexity. 
Thus, Tang et al.~\cite{tang2016privacy} apply homomorphic encryption on the rating data of a target users' friends only, i.e., a small subset of users, to improve computational efficiency.
Besides homomorphic encryption, \emph{federated learning}~\cite{mcmahan2017communication} is used to lower users' privacy risks~\cite{lin2020meta,han2021deeprec,10.1145/3506715,perifanis2022federated}. 
Specifically, instead of a user's rating data, the parameters of the user's local recommendation model are utilized in the recommendation process.
For example, Perifanis and Efraimidis~\cite{perifanis2022federated} combine federated learning with neural collaborative filtering~\cite{he2017neural} to improve privacy.
However, since federated learning could still leak user data~\cite{nasr2019comprehensive,10.1145/3510032}, research proposes to learn a user's local model by utilizing only a subset of the rating data~\cite{anelli2021put,robustnessofmetamf,chen2022proactively}.
Moreover, \emph{differential privacy (DP)}~\cite{dwork2014algorithmic,dwork2008differential} has been leveraged for collaborative filtering recommender systems~\cite{10.1145/3501812,10.1145/3394138,10.1145/3506715,gao2020dplcf,chen2022differential,zhu2013differential}. 
These techniques add randomness to users’ data to hide the actual data. Therefore, they face a trade-off between accuracy and privacy 
(e.g.,~\cite{berkovsky2012impact}). 
To address this trade-off, Xin and Jaakkola~\cite{10.5555/2969033.2969119} assume a moderate number of public users who tolerate disclosing their rating data.
With this unprotected rating data, recommendation accuracy can be preserved while the privacy requirements of the remaining users are respected. 

It has been shown in several studies~\cite{kowald2020unfairness,mansoury2020feedback,abdollahpouri2019unfairness} that users often receive more recommendations for popular items, and correspondingly non-popular items receive less exposure. 
This behavior of recommender systems, which is referred to as popularity bias, leads to disparate, i.e., unfair, treatment of less popular items.
Dwork et al.~\cite{dwork2012fairness} and Zemel et al.~\cite{zemel2013learning} show that, formally, there is a close connection between fairness and DP.
However, the sole application of DP is insufficient to ensure fairness due to correlations within the dataset~\cite{ekstrand2018privacy}.
Moreover, Ekstrand et al.~\cite{ekstrand2018privacy} and Agarwal~\cite{agarwal2020trade} highlight a trade-off between user privacy and fairness.
Overall, related work suggests that DP can severely impact recommendations in different ways, for example, result in popularity bias.
Therefore, we believe that it is important to evaluate the proposed approach, i.e., \emph{ReuseKNN}, also in terms of item popularity bias.



Similar to our work, previous research by Zhu et al.~\cite{zhu2013differential} prevents the inference of neighbors' rating data by applying DP to the users' rating data in \emph{UserKNN}. However, to preserve recommendation accuracy, Zhu et al. vary the degree of randomness that is added to all users' rating data based on the sensitivity of the data.
In contrast, \emph{ReuseKNN} preserves recommendation accuracy by adding randomness only where it is necessary, i.e., to vulnerable users with a high privacy risk. 
In the remainder of the paper, we use a variant of Zhu et al.'s approach that is comparable to the proposed approach as baseline (i.e., \emph{UserKNN}$^{full}_{DP}$) for our experiments.


\section{Problem Definition}
\label{sec:problem}
In the following, 
we discuss one key vulnerability of \emph{UserKNN}, that poses privacy risks to the neighbors utilized in the recommendation process.
Also, we precisely model the adversary's goal, i.e., the inference of the neighbors' rating data.
Moreover, a summary of the notation used in this paper is given in Table~\ref{tab:notation}.

\subsection{Vulnerability Analysis of UserKNN}
\label{subsec:userknn}
Typically, a user-based \emph{KNN} recommender system $\mathcal{R}^k$, i.e., \emph{UserKNN}, generates an estimated rating score for a rating query  of a target user $u$ and a target item $i$ by utilizing the ratings of $k$ other users that have rated $i$, i.e., the $k$ nearest neighbors $N^k_{u, i}$:
\begin{equation}
   \mathcal{R}^k(u, i) = \frac{\sum_{n \in N^k_{u, i}} sim(u, n) \cdot r_{n, i}}{\sum_{n \in N^k_{u, i}} sim(u, n)}
   \label{eq:userknn}
\end{equation}
where $sim(u, n)$ is the similarity between target user $u$ and neighbor~$n$, commonly determined via Pearson's correlation coefficient~\cite{benesty2009pearson} or Cosine similarity between the users' rating vectors.
For \emph{UserKNN}, the neighborhood $N^k_{u, i}$ used for generating recommendations for target user $u$ and item $i$, comprises the $k$ most similar neighbors:
\begin{equation}
    N^k_{u, i} =\ \stackrel{k}{\argmax_{c \in U_i}} sim(u, c)
    \label{eq:userknn_neighborhood}
\end{equation}
where $U_i$ are all users that have rated $i$ and $sim$ is the similarity metric. 
\emph{UserKNN} utilizes the rating data of the target user's $k$ nearest neighbors to generate an estimated rating score (see Equation~\ref{eq:userknn}).
Therefore, the estimated rating score $\mathcal{R}^k(u, i)$ for target user $u$ and item $i$ is linked to the neighbors' rating data. 
Through learning the behavior of \emph{UserKNN}, 
the estimated rating score 
could reveal the rating data of users that have been used as neighbors~\cite{calandrino2011you}. 
Therefore, the privacy threat for users can be traced back to them being utilized as neighbors in the recommendation process.

\subsection{Attack Model}
In this work, we assume that a user with malicious intent, i.e., the \emph{adversary} $a$, exploits the vulnerability above via querying estimated rating scores from the recommender system
, i.e., $\mathcal{R}^k(a) = \{\mathcal{R}^k(a, i_1), \mathcal{R}^k(a, i_2), \dots, \mathcal{R}^k(a, i_l)\}$, where $\mathcal{R}^k(a, i_j)$ is the estimated rating score for item $i_j \in Q_a$ and $Q_a$ is the set of $a$'s queries. 
The adversary $a$ can target a specific user $n$ by increasing the likelihood of $n$ being used as neighbor.
To achieve this, $a$ would modify its own user profile $R_a$ such that it (partially) matches $n$'s profile.
Moreover, $a$ can exploit publicly available data $P$, e.g., public rating data, product reviews, tweets, or lists of similar items, to better learn the behavior of \emph{UserKNN}~\cite{calandrino2011you}.
Given these assumptions, the adversary aims to infer the rating data of a neighbor $n$ used to generate the estimated rating scores:
\begin{equation}
    \label{eq:attack}
    Pr[r_{n, i_1}, r_{n, i_2}, \dots, r_{n, i_l} | \mathcal{R}^k(a, i_1), \mathcal{R}^k(a, i_2), \dots, \mathcal{R}^k(a, i_l), P \cup R_a]
\end{equation}
where $r_{n, i_j}$ is the rating score of neighbor $n$ for item $i_j$.
Note that if a user is used as neighbor for many rating queries, many ratings could be targeted by an adversary.
Thus, the degree to which a user's rating data is used in the recommendation process is an important indicator of this user's privacy risk (see the $\mathrm{DataUsage}@k$ metric in Section~\ref{subsec:privacy_risk}).

{Given this attack model, the privacy threat lies on the rating-level, i.e., the inference of neighbors' rating scores, and therefore, our approach aims at protecting the neighbors' rating scores. 
In the remainder of this work, we evaluate our approach in a rating-prediction task, since this fits well to our problem statement above (see Appendix~\ref{sec:a_topn} for results of a ranking-based experiment).}

\begin{table}[!t]
    \centering
    \caption{Overview of the notation used in this paper.}   
    \begin{footnotesize}
    \begin{tabular}{l l l} \toprule
    Symbol & Description \\ \midrule
    $k$ & Number of neighbors to process a rating query for target user $u$ and target item $i$.\\
    $Q_u$ & Rating queries for target user $u$, i.e., the items in $u$'s test set $R^{test}_u$.\\
    $\mathcal{R}^k$ & User-based \emph{KNN} recommender system utilizing $k$ neighbors to predict ratings.\\
    $\mathcal{R}^k(u, i)$ & Estimated rating score for target user $u$ and target item $i$ by recommender system $\mathcal{R}^k$.\\
    $\mathcal{R}^k_{top}(u)$ & Items with the highest estimated rating score for target user $u$. \\
    $r_{u, i}$ & Rating score of user $u$ to item $i$.\\
    $U$ & The set of users. \\
    $U_i$ & The set of users that rated item $i$. \\
    $I$ & The set of items. \\
    $I_u$ & The set of items rated by user $u$. \\
    $R$ & The set of ratings. \\
    $N^k_{u, i}$ & The $k$ nearest neighbors for target user $u$ and target item $i$.\\
    $N_{u, i}$ & Neighbors of target user $u$ and rated item $i$.\\
    $N_u$ & The set of neighbors for target user $u$ across all rating queries.\\
    {$N_u^{(q)}$} & The set of neighbors for target user $u$ across $q$ rating queries.\\
    $sim(u, n)$ & Similarity score between target user $u$ and neighbor~$n$. \\
    $reusability(c|u)$ & Reusability score of candidate neighbor $c$ for target user $u$. \\
    $ranking(\cdot)$ & The ranking function that ranks candidate neighbors w.r.t. similarity and reusability.\\
    $\tau$ & Data usage threshold, i.e., the maximal usage of a user's data that is tolerated. \\
    $m_{DP}$ & Differentially privacy mechanism that utilizes plausible deniability.\\
    $\epsilon$ & Privacy parameter. \\
    $S$ & Secure users that do not need to be protected with DP.\\
    $V$ & Vulnerable users that need to be protected with DP.\\
    $R_S$ & Rating data of secure users.\\
    $\Tilde{R}_V$ & DP-protected rating data of vulnerable users.\\
    $\alpha$ & Significance level used for the statistical tests.\\
    $\sigma_x$ & Sample standard deviation of variable $x$.\\
    $\sigma_{x,y}$ & Sample covariance of variables $x$ and $y$.\\
    \bottomrule
    \end{tabular}
    \end{footnotesize}
    \label{tab:notation}
\end{table}
\section{Approach}
In the following, we first schematically illustrate \emph{UserKNN}'s and \emph{ReuseKNN}'s recommendation process based on an illustrative example.
Then, we outline the two neighborhood reuse strategies of the \emph{ReuseKNN} recommender system (Section~\ref{subsec:reuseknn}).
Finally, we present \emph{ReuseKNN}$_{DP}$, i.e., neighborhood reuse with \emph{differential privacy (DP)} (Section~\ref{subsec:reuseknn_dp}).
\subsection{Example of the Recommendation Process in \emph{UserKNN} and \emph{ReuseKNN}}
\label{subsec:example}
\begin{figure}[!t]
    \centering
    \includegraphics[width=0.9\linewidth]{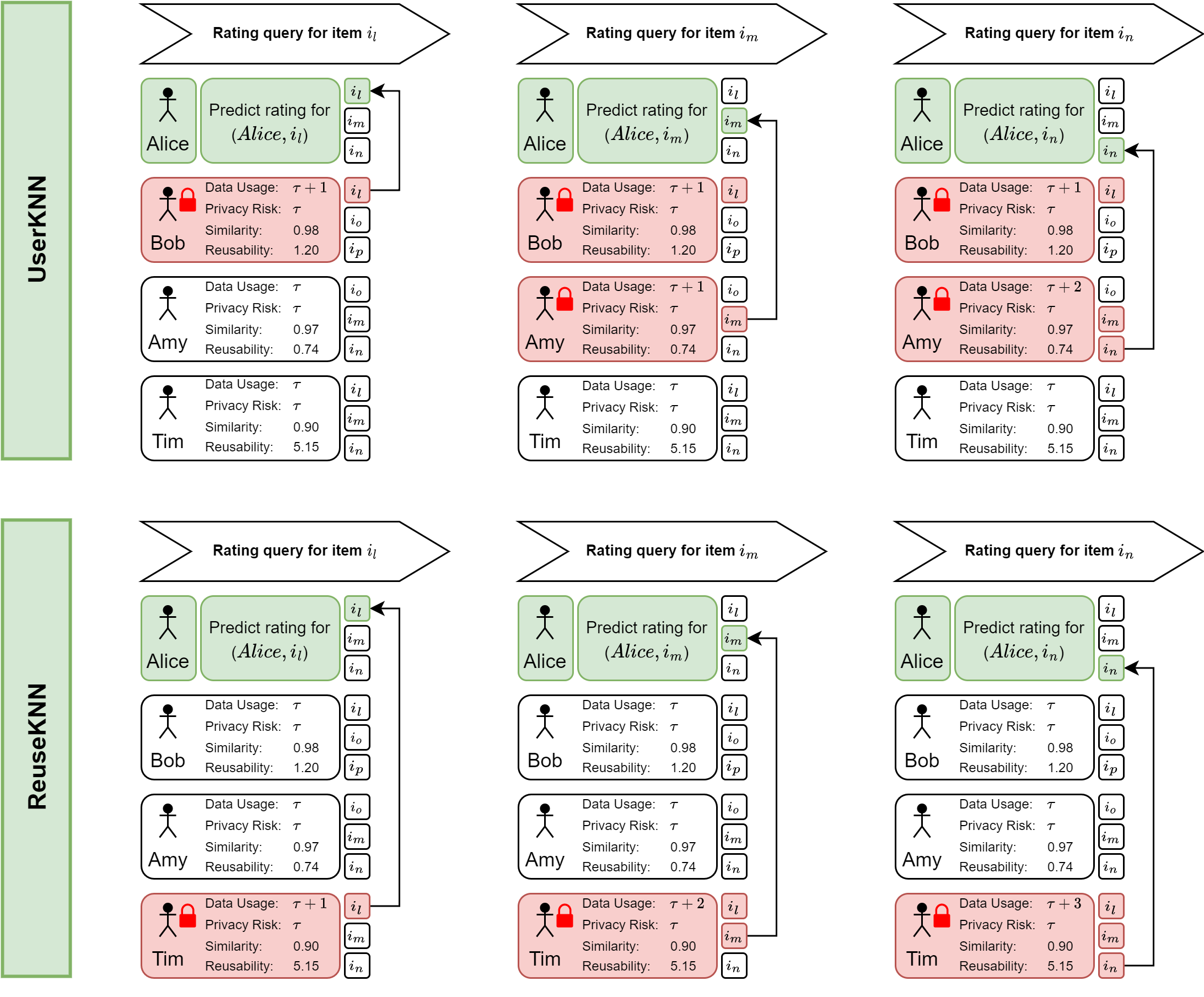}
    \caption{Schematic illustration of the recommendation process for three rating queries in Alice's query set $Q_{Alice}$ for \emph{UserKNN} and \emph{ReuseKNN}. 
    Furthermore, a green shaded item indicates that the rating score for this item is estimated for the target user and a red shaded item indicates that the rating score of a neighbor has been utilized for the rating estimation.
    Traditional \emph{UserKNN} selects those users as neighbors, that rated the queried item and have the highest similarity-value; in this toy example, those are Bob and Amy. 
    Thus, Bob and Amy are vulnerable and need to be protected with DP.
    In contrast, \emph{ReuseKNN} utilizes Tim as neighbor. 
    As such, \emph{ReuseKNN} substantially increases reusability (5.15 instead of 1.2 and 0.74), at the price of a slightly reduced similarity (0.90 instead of 0.98 and 0.97).
    This way, only Tim is vulnerable and is the only neighbor that needs to be protected with DP, as Bob and Amy remain unused.}
    \label{fig:overview}
\end{figure}

Figure~\ref{fig:overview} provides a schematic illustration of \emph{UserKNN}'s and \emph{ReuseKNN}'s recommendation process, showing the interplay between a user's data usage and their privacy risk.
For simplicity, we assume that Bob, Amy, and Tim have been used as neighbors for $\tau$ rating queries, i.e., data usage and privacy risk is $\tau$.
To process Alice's rating queries for items $i_l$ and $i_m$, \emph{UserKNN} selects Bob and Amy as neighbors, as they have the highest similarity-values across all users that rated the queried items.
Due to the usage of Bob's and Amy's data, their data usage exceeds threshold $\tau$ and DP needs to be applied.
For the rating query for item $i_n$, again, Amy is utilized in the recommendation process.
Since she is already protected with DP, her privacy risk remains at $\tau$.
This is different to how \emph{ReuseKNN} processes rating queries.
For the rating queries for items $i_l$, $i_m$, and $i_n$, \emph{ReuseKNN} selects Tim as neighbor, as Tim has a substantially higher reusability-value and only marginally smaller similarity than Bob and Amy.
Therefore, only Tim's data usage exceeds $\tau$, and DP is needed to protect Tim.

In summary, in this illustrative example, \emph{UserKNN} leads to two vulnerable users, i.e., Bob and Amy, that need to be protected with DP.
In contrast, \emph{ReuseKNN} leads to only one vulnerable user, i.e., Tim, to which DP has to be applied.

\subsection{ReuseKNN}
\label{subsec:reuseknn}
The key feature of \emph{ReuseKNN} is to reuse neighbors from a target user $u$'s previous rating queries to minimize the cardinality of the neighborhood $N_u = \bigcup_{i \in Q_u} N^k_{u, i}$ across all rating queries $Q_u$.
As illustrated in Figure~\ref{fig:approach}, this means that \emph{ReuseKNN} decreases the data usage for most users, i.e., secure users, and in this way, also their privacy risk.
Plus, \emph{ReuseKNN} decreases the number of highly reused neighbors, i.e., vulnerable users with high data utilization and thus, high privacy risk.

In addition to the similarity, \emph{ReuseKNN} also considers the extent to which a target user $u$ could reuse candidate neighbor $c$ as a neighbor for many rating queries, i.e., $reusability(c|u)$.
Since both, similarity and reusability scores are differently distributed across their respective numeric ranges, we rank candidate neighbors according to their scores.
Formally, for a user $u$, the rank $ranking(u) = |\{v \in U\setminus\{u\}: f(v) \leq f(u)  \}|$, where $U$ is the set of all users and $f$ measures the $similarity$ or $reusability$ score.
Note that $ranking(u) > ranking(v)$ if $f(u) > f(v)$ for users $u$ and $v$, and that $ranking(u) = ranking(v)$ in case $f(u) = f(v)$.
With this, the $k$ neighbors $N^k_{u, i}$ are selected based on similarity and also, reusability. Formally: 
\begin{equation}
    \label{eq:tradeoff}
    N^k_{u, i} =\ \stackrel{k}{\argmax_{c \in U_i}} \big[ranking(sim(u, c)) + ranking(reusability(c|u))\big]
\end{equation}
where $U_i$ are all users that rated item $i$, $sim$ measures the similarity between two users, and $reusability$ depends on the given neighborhood reuse strategy of \emph{ReuseKNN}. 
In case multiple candidate neighbors have equal values for $ranking(sim(u, c)) + ranking(reusability(c|u))$, we choose these neighbors at random.

To estimate a candidate neighbor's $reusability$ score, \emph{ReuseKNN} utilizes two neighborhood reuse strategies: \emph{Expect} and \emph{Gain}.
The unpersonalized \emph{Expect} strategy measures a candidate neighbor's reusability for an average target user, whereas the personalized \emph{Gain} strategy measures the reusability for a specific target user.
Next, we discuss two strategies to increase the reusability of a target user's neighbors, i.e., unpersonalized and personalized neighborhood reuse.

\vspace{2mm} \noindent \emph{Unpersonalized Neighborhood Reuse: Expect.}
The more users rated an item, the more likely it is that a random target user will query a rating estimation for this item.
Following this intuition, \emph{Expect} promotes candidate neighbors that rated many popular items and penalizes candidate neighbors that either rated only a few items or many unpopular items. 
For \emph{Expect}, the reusability score of candidate neighbor $c$ is defined by:
\begin{equation}
    reusability(c|u) = reusability(c) = \sum_{i \in I_c} \frac{|U_i|}{|U|} \label{eq:popularity}
\end{equation}
where $u$ is the target user, $I_c$ are the items $c$ rated, $U_i$ are the users that rated an item $i$, and $U$ is the set of all users.
In this case, $reusability(c)$ is the summed-up popularity of $c$'s rated items and measures the \emph{expected} number of a random user's rating queries for which $c$ could be used as a neighbor.
This means that the reusability of a candidate neighbor is estimated for an average user and not for a specific target user (i.e., unpersonalized). 

\vspace{2mm} \noindent \emph{Personalized Neighborhood Reuse: Gain.}
In contrast to unpersonalized neighborhood reuse, \emph{Gain} measures a candidate neighbor's reusability for a specific target user. 
In detail, \emph{Gain} quantifies how many of a target user's ratings a candidate neighbor could have covered in the past, i.e., how many ratings the target user could have \emph{gained} from the candidate neighbor.
Thus, \emph{Gain} gives the fraction of a target user $u$'s rated items, for which a candidate neighbor $c$ could have served as a neighbor:
\begin{equation}
    reusability(c|u) = \frac{|I_u \cap I_c|}{|I_u|} \label{eq:gain}
\end{equation}
where $I_u$ are the items rated by $u$ and $I_c$ are the items rated by $c$.
In contrast to the unpersonalized \emph{Expect} strategy, \emph{Gain}'s reusability score depends on a specific target user (i.e., personalized).

\subsection{ReuseKNN$_{DP}$: Neighborhood Reuse and Differential Privacy}
\label{subsec:reuseknn_dp}
\emph{ReuseKNN} leads to a minimal number of highly reused neighbors, i.e., vulnerable users, who are utilized more often as neighbors than the data usage threshold $\tau$ would allow.
\emph{ReuseKNN}$_{DP}$ addresses this high privacy risk resulting from the frequent usage of vulnerable users (see Section~\ref{sec:problem}) by adding DP to our neighborhood reuse strategies. 
Specifically, for a rating query for user $u$ and item $i$, a privacy mechanism $m_{DP}$ is applied to the ratings for $i$ of vulnerable users $V$ that are used as neighbors, i.e., $\Tilde{R}_V = \{m_{DP}(r_{n, i}): n \in N^k_{u, i} \cap V\}$.
In this way
, \emph{ReuseKNN}$_{DP}$ utilizes real ratings of secure users $S$, i.e., $R_S = \{r_{n, i}: n \in N^k_{u, i} \cap S\}$, plus the modified ratings $\Tilde{R}_V$ of vulnerable users, to generate the estimated rating score $\mathcal{R}^k(u, i)$:
\begin{equation}
    \mathcal{R}^k(u, i) = \frac{\sum_{n \in N^k_{u, i} \cap S} sim(u, n) \cdot r_{n, i} + \sum_{n \in N^k_{u, i} \cap V} sim(u, n) \cdot m_{DP}(r_{n, i})}{\sum_{n \in N^k_{u, i}} sim(u, n)}
    \label{eq:reuseknn_dp}
\end{equation}
In detail, the privacy mechanism $m_{DP}$ utilizes \emph{randomized responses}~\cite{warner1965randomized} to achieve DP~\cite{dwork2014algorithmic}.
With this, intuitively, neighbors can plausibly deny that their real rating was used in the recommendation process.
The privacy mechanism $m_{DP}$ flips a fair coin and if the coin is heads, the vulnerable neighbor's real rating is utilized in the recommendation process.
If the coin is tails, $m_{DP}$ flips a second fair coin to decide whether to utilize the vulnerable neighbor's real rating or a random rating drawn from a uniform distribution over the range of ratings.
With this, the adversary is unaware whether the utilized rating is real, or random, which leads to the privacy-guarantees within the DP-framework~\cite{dwork2014algorithmic}.
Specifically, 
\begin{equation}
    \frac{Pr[\text{Adversary's assumption: Real rating | Truth: Real rating}]}{Pr[\text{Adversary's assumption: Real rating | Truth: Random rating}]} = \frac{0.75}{0.25} = 3 \leq e^\epsilon
\end{equation}
which results in a privacy parameter of $\epsilon = \ln 3$.
Reconsidering user-based \emph{KNN}'s vulnerablility (see Equation~\ref{eq:userknn}), this means that if a neighbor $n$ is considered as vulnerable, the DP-protected rating is used in the recommendation process instead of the real rating for item $i$ (see Equation~\ref{eq:reuseknn_dp}).
This impacts the adversary $a$'s objective (see Equation~\ref{eq:attack}) of inferring $n$'s rating data given the estimated rating scores for which $n$ was used as neighbor and its own rating data $R_a$ in combination with public knowledge $P$ (see Section~\ref{sec:problem}).
Since a maximum of $\tau$ (i.e. the data usage threshold) real ratings of $n$ are used by the recommender system, the remaining ratings are DP-protected.
Thus, the adversary is not aware whether the inferred rating data is the original rating data, or is random rating data as generated by the $m_{DP}$ mechanism:
\begin{equation}
     Pr[r_{n, i_1}, \dots, r_{n, i_\tau}, m_{DP}(r_{n, i_{\tau+1}}), \dots, m_{DP}(r_{n, i_l}) | \mathcal{R}^k(a, i_1), \mathcal{R}^k(a, i_2), \dots, \mathcal{R}^k(a, i_l), P \cup R_a]
\end{equation}
where $r_{n, i_j}$ is $n$'s rating for item $i_j$ and $\mathcal{R}^k(a, i_j)$ is the estimated rating score of $i_j$ for adversary $a$.
Through combining non-DP and DP-ratings, \emph{ReuseKNN}$_{DP}$ yields the following privacy parameter $\epsilon$ for each of a vulnerable user's, in this case $n$, utilized ratings (for details see Appendix~\ref{sec:a_privacyanalysis}):
\begin{equation}
    \epsilon = \ln \Bigg(3 + 4 \cdot \frac{Pr[\text{Non-DP rating}]}{Pr[\text{DP rating}]}\Bigg)
\end{equation}
In this way, \emph{ReuseKNN}$_{DP}$ combines neighborhood reuse with DP to reduce the number of neighbors to which DP needs to be applied and to ensure privacy.
Overall, \emph{ReuseKNN}$_{DP}$ can use two neighborhood reuse strategies with DP (for details see Section~\ref{subsec:reuseknn}): 
\begin{enumerate}
    \item \emph{Expect}$_{DP}$: Unpersonalized neighborhood reuse combined with DP
    \item \emph{Gain$_{DP}$}: Personalized neighborhood reuse combined with DP
\end{enumerate}

\section{Experimental Setup} \label{sec:expsetup}
We utilize a two-stage evaluation procedure to compare and evaluate the two neighborhood reuse strategies of (i) \emph{ReuseKNN} and (ii) \emph{ReuseKNN}$_{DP}$:

\vspace{2mm} \noindent \emph{Neighborhood Reuse without DP: ReuseKNN.}
In the first stage, we evaluate \emph{ReuseKNN} without protecting vulnerable neighbors with DP in order to better understand the advantages and disadvantages of the proposed neighborhood reuse strategies. 
Hence, we compare \emph{Expect} and \emph{Gain} to distill the impact of neighborhood reuse for recommendations. 

\vspace{2mm} \noindent \emph{Neighborhood Reuse with DP: ReuseKNN$_{DP}$.}
In the second stage, we combine \emph{ReuseKNN} with DP to protect vulnerable users, i.e., \emph{ReuseKNN}$_{DP}$.
We compare our neighborhood reuse strategies \emph{Expect}$_{DP}$ and \emph{Gain}$_{DP}$ to investigate how \emph{ReuseKNN}$_{DP}$ can address the accuracy-privacy trade-off.


\subsection{Baselines}
\label{subsec:baselines}
We compare $\emph{ReuseKNN}$ and \emph{ReuseKNN}$_{DP}$ to five different KNN-based baselines. 
Concretely, for \emph{ReuseKNN}, i.e., neighborhood reuse without DP, we use two non-DP baselines:
\begin{enumerate}
    \item \emph{UserKNN}~\cite{herlocker1999algorithmic}: Traditional \emph{UserKNN} without neighborhood reuse. No users are protected with DP (Vulnerable users $V=\emptyset$).
    \item \emph{UserKNN+Reuse}: A variant of \emph{UserKNN} with neighborhood reuse. Initially, for the first rating query, e.g., for item $j$, the $k$ most similar users that rated $j$ are selected as neighbors as in case of \emph{UserKNN}. 
    However, for the following rating queries, e.g., for item $i$ and user $u$, $k^{prev} = \min\{k, |N_{u, i}|\}$ neighbors from all previous rating queries that rated $i$ (i.e., $N_{u, i}$) are reused.
    If too few previous neighbors rated $i$, i.e., $k^{prev} < k$, a minimal set of $k^{new} = k - k^{prev}$ new neighbors is additionally used, as given by: 
    \begin{equation}
        N^k_{u, i} = \stackrel{k^{prev}}{\argmax_{n \in N_{u, i}}} sim(u, c)\ \cup \stackrel{k^{new}}{\argmax_{c \in U_i\setminus N_{u, i}}} sim(u, c)
    \end{equation}
    where $U_i$ are all users that rated item $i$. 
    Similar to \emph{UserKNN}, \emph{UserKNN+Reuse} assumes that no users are vulnerable ($V = \emptyset$). Thus, no users are protected with DP.
\end{enumerate}
For \emph{ReuseKNN}$_{DP}$, i.e., neighborhood reuse with DP, we use three DP baselines:

\begin{enumerate}
    \item \textit{UserKNN}$_{DP}$: 
    A variant of \emph{UserKNN}, but DP is applied to vulnerable users $V = \{u \in U: \mathrm{DataUsage}@k(u) > \tau\}$. Please see Section~\ref{s:parameters} for the exact $\tau$ values.
    \item \textit{UserKNN+Reuse}$_{DP}$: A variant of \textit{UserKNN+Reuse}, but DP is applied to vulnerable users $V = \{u \in U: \mathrm{DataUsage}@k(u) > \tau\}$. Please see Section~\ref{s:parameters} for the exact $\tau$ values.
    \item \textit{UserKNN}$^{full}_{DP}$: Traditional differentially-private \emph{UserKNN} where DP is applied to the full set of users, i.e., $V = \{u \in U: \mathrm{DataUsage}@k(u) \geq 0\}$ (similar to the rating perturbation in~\cite{zhu2013differential}).
\end{enumerate}
To evaluate \emph{ReuseKNN}$_{DP}$, we use the three DP baselines, as well as non-DP \emph{UserKNN}.
With this, we can compare \emph{ReuseKNN}$_{DP}$ to two contrastive baselines: \emph{UserKNN}$^{full}_{DP}$, which protects all users with DP, and \emph{UserKNN}, which does not apply DP at all.

\subsection{Evaluation Metrics}

\begin{table}[!t]
    \centering
    \caption{Overview of the seven evaluation metrics used in this work. $\searrow$ indicates that lower values are better and $\nearrow$ indicates that higher values are better. $q$ is the number of queries and $k$ is the number of neighbors. With $\bullet$, we indicate the evaluation stage in which the metric is used.}
    \begin{footnotesize}
    \begin{tabular}{l @{\hspace{.99\tabcolsep}} l @{\hspace{.99\tabcolsep}} c @{\hspace{.99\tabcolsep}} l c @{\hspace{.99\tabcolsep}} c} \toprule
    & & & & \multicolumn{2}{c}{Evaluation Stage} \\ 
    \cmidrule(lr){5-6}
    Evaluation Criterion & Evaluation Metric & Objective & Short description & \emph{ReuseKNN} & \emph{ReuseKNN}$_{DP}$ \\ \midrule
    \multirow{2}{*}{Neighborhood Reuse} & $\mathrm{Neighbors}@q$ & $\searrow$ & Neighborhood size & $\bullet$ &  \\
    & $\mathrm{CoRatings}@q$ & $\nearrow$ & No. of co-rated items & $\bullet$ & \\ \midrule
    Accuracy & $\mathrm{MAE}@k$ & $\searrow$ & Mean absolute error & $\bullet$ & $\bullet$ \\ \midrule
    \multirow{2}{*}{Privacy} & $|V|$ & $\searrow$ & Percentage of vulnerable users & $\bullet$ &  \\
    & $\mathrm{PrivacyRisk}@k$ & $\searrow$ & Privacy risk of users &  & $\bullet$ \\ \midrule
    \multirow{2}{*}{Popularity Bias} & $\text{PP-Corr}@k$ & $\searrow$ & Positivity-popularity correlation &  & $\bullet$ \\
    & $\mathrm{Coverage}@k$ & $\nearrow$ & Percentage of item coverage &  & $\bullet$ \\ \bottomrule
    \end{tabular}
    \end{footnotesize}
    \label{tab:evaluation_metrics}
\end{table}

We test the proposed approach in two evaluation stages using the following evaluation criteria and metrics (see Table~\ref{tab:evaluation_metrics} for an overview):

\subsubsection{Neighborhood Reuse}

To evaluate the degree to which \emph{ReuseKNN} can reuse neighbors from previous rating queries, we measure the size of a target user's neighborhood after multiple queries.
Plus, we study if the reused neighborhoods are capable of generating meaningful recommendations via measuring the number of co-rated items between the neighborhood and the target user.

\vspace{2mm} \noindent \emph{Neighborhood Size.} 
For every rating query of a target user $u$, $k$ neighbors are required to generate the recommendation.
In the worst case, no neighbors from previous rating queries can be reused. Thus, after $q$ queries, $|N_u|=\min\{ q \cdot k, |U|-1 \}$ for $U$ being the set of all users.
In the best case, $u$ reuses the same $k$ neighbors for all $q$ queries, i.e., $|N_u| = k$.
To quantify how many of $u$'s neighbors are reused, we measure the size of $u$'s neighborhood after $q$ rating queries:
\begin{equation}
    \mathrm{Neighbors}@q(u) = |N^{(q)}_u|
\end{equation}
where $N^{(q)}_u$ is $u$'s set of neighbors after $q$ rating queries.
With that, we test how well our neighborhood reuse strategies of \emph{ReuseKNN}, i.e., neighborhood reuse only, can reuse a target user's neighbors for multiple rating queries.

\vspace{2mm} \noindent \emph{Number of Co-Ratings.} 
The utilization of fewer neighbors across many rating queries might impact the accuracy of recommendations.
Therefore, we test if a target user's neighbors are beneficial for recommendation accuracy, i.e., ``reliable''.
One important characteristic of these reliable neighbors is the number of co-rated items with the target user~\cite{desrosiers2010comprehensive,adomavicius2012impact}.
Thus, we measure the average number of co-rated items between a target user $u$ and its neighbors $n \in N_u$ after $q$ rating queries:
\begin{equation}
    \mathrm{CoRatings}@q(u) = \frac{1}{|N_u^{(q)}|}\sum_{n \in N_u^{(q)}} |I_u \cap I_n|
\end{equation}
where $I_u$ are the items rated by target user $u$ and $I_n$ are the items rated by neighbor $n$.
With this, we test how beneficial the neighborhoods are for generating accurate recommendations.

\subsubsection{Accuracy}
To quantify the accuracy of a target user's recommendations, we rely on the widely-used mean absolute error metric (MAE). 
{We use MAE to measure how accurate the rating scores can be predicted, because of the way in which we apply DP, i.e., via adding noise to the neighbors' rating values in order to protect against the disclosure of these ratings (see Section~\ref{sec:problem}).}
According to Herlocker et al.~\cite{herlocker1999algorithmic} the number of neighbors $k$ has an impact on the recommendation accuracy.
Thus, we test the accuracy of $u$'s recommendations for $k \in \{5, 10, 15, 20, 25, 30\}$.
Therefore, $\mathrm{MAE}@k(u)$ quantifies the accuracy of $u$'s recommendations when $k$ neighbors are used to generate a recommendation.
More formally:  
\begin{equation}
    \mathrm{MAE}@k(u) = \frac{1}{|R^{test}_u|} \sum_{r_{u, i} \in R^{test}_u} |r_{u, i} - \mathcal{R}^k(u, i)| 
\end{equation}
where the predicted rating score $\mathcal{R}^k(u, i)$ is compared to the real rating scores $r_{u, i} \in R^{test}_u$ in $u$'s test set.
We note that the items for which $R^{test}_u$ comprises ratings are the ones that are in $u$'s set of rating queries $Q_u$.
We use the $\mathrm{MAE}@k(u)$ metric for evaluating both, \emph{ReuseKNN}, i.e., neighborhood reuse only, and \emph{ReuseKNN}$_{DP}$, i.e., neighborhood reuse with DP.

\subsubsection{Privacy}
\label{subsec:privacy_risk}
Liu and Terzi~\cite{liu2010framework} provide a framework to measure a user's privacy risk in computational systems such as recommender systems based on the visibility of the user's data. 
In our work, we relate this visibility to how often a user's rating data was utilized in the recommendation process.
As such, the $\mathrm{DataUsage}@k(u)$ metric counts for how many rating queries a user $u$ was used as a neighbor.
Similar to $\mathrm{MAE}@k(u)$, we also relate the usage of $u$'s data to the number of neighbors $k$ used to generate recommendations. Formally: 
\begin{equation}
    \mathrm{DataUsage}@k(u) = \sum_{v \in U} \sum_{i \in Q_v} \mathbbm{1}_{N^k_{v, i}}(u)\label{eq:DataUsage}
\end{equation}
where $U$ is the set of all users, $Q_v$ is the set of items for which user $v$ queries estimated ratings, and $\mathbbm{1}_{N_{v, i}}(u)$ is the indicator function of user $u$ being in $v$'s set of neighbors $N_{v, i}$ for an item $i$.

\vspace{2mm} \noindent \emph{Percentage of Vulnerable Users.} 
As mentioned earlier, the main goal of neighborhood reuse is to reduce the number of users that need to be protected with DP. 
The $\mathrm{DataUsage}@k$ definition allows us to identify these vulnerable users $V$, i.e., the set of users whose data is utilized more often than the adjustable privacy risk threshold $\tau$ allows:  
\begin{equation}
    V = \{u \in U: \mathrm{DataUsage}@k(u) > \tau \}
\end{equation}
where $U$ is the set of all users.
Thus, the percentage of vulnerable users relates to what fraction of users DP has to be applied to (i.e., $|V| / |U|$).
We use this metric to evaluate \emph{ReuseKNN}, i.e., neighborhood reuse only.

\vspace{2mm} \noindent \emph{Privacy Risk.} 
We apply DP to a user $u$'s data if $\mathrm{DataUsage}@k(u) > \tau$. 
This way, only the first $\tau$ utilized ratings contribute to $u$'s privacy risk, since for the remaining ratings that are utilized, privacy is guaranteed via the DP-framework (see Section~\ref{subsec:reuseknn_dp}):
\begin{equation}
    \mathrm{PrivacyRisk}@k(u) = \min [\tau, \mathrm{DataUsage}@k(u) ]
\end{equation}
We use $\mathrm{PrivacyRisk}@k$ to measure the users' privacy risk when neighborhood reuse is combined with DP, i.e., \emph{ReuseKNN}$_{DP}$.

\subsubsection{Item Popularity Bias}
One might be concerned that neighborhood reuse could lead to exploiting users as neighbors that rated many popular items, which could result in more positive estimated rating scores for popular items.
To test for this item popularity bias, we analyze all items for which the recommender system estimates high rating scores, i.e., ``top items''.
For a recommender system model $\mathcal{R}$ and $k$ neighbors, a user $u$'s set of top items is given by:
    \smash{$\mathcal{R}^k_{top}(u) =\ \stackrel{n}{\argmax_{i \in Q_u}} \mathcal{R}^k(u, i)$}
, where $Q_u$ are the items in $u$'s query set. In our case, we set $n=10$.

\vspace{2mm} \noindent \emph{Positivity-Popularity Correlation.}
To study if higher estimated rating scores are given to popular items, we follow Kowald et al.~\cite{kowald2020unfairness} and correlate an item's popularity with its occurrences in users' sets of top items: 
\smash{$\mathrm{ItemFreq}^+@k(i) = \sum_{u \in U} \mathbbm{1}_{\mathcal{R}^k_{top}(u)}(i)$}
where \smash{$\mathbbm{1}_{\mathcal{R}^k_{top}(u)}(i)$} indicates whether item $i$ is in user $u$'s  set of top items \smash{$\mathcal{R}^k_{top}(u)$}.
Plus, an item $i$'s popularity is given by: $\mathrm{ItemPop(i)} = |U_i| / |U|$, where $U$ is the set of all users and $U_i$ are the users that rated $i$.
We compute the Pearson correlation coefficient~\cite{benesty2009pearson} between the two variables $\mathrm{ItemFreq}^+$ and $\mathrm{ItemPop}$ to identify item popularity bias:
\begin{equation}
    \text{PP-Corr}@k = \frac{\sigma_{\mathrm{ItemFreq}^+@k, \mathrm{ItemPop}@k}}{\sigma_{\mathrm{ItemFreq}^+@k} \cdot \sigma_{\mathrm{ItemPop}@k}}
\end{equation}
where $\sigma_{\mathrm{ItemFreq}^+@k, \mathrm{ItemPop}@k}$ is the sample covariance between $\mathrm{ItemFreq}^+@k$ and $\mathrm{ItemPop}@k$.
Furthermore, the sample standard deviations are given by $\sigma_{\mathrm{ItemFreq}^+@k}$ and $\sigma_{\mathrm{ItemPop}@k}$.

\vspace{2mm} \noindent \emph{Item Coverage.} 
In addition to evaluating the correlation between an item's estimated rating score and its popularity, we measure the fraction of items that are a top item for at least one user.
For this, we use the Item Coverage metric~\cite{evaluatingherlocker} given by:
\begin{equation}
    \mathrm{Coverage}@k = \frac{1}{|I|} \; \left|\bigcup_{u \in U} \mathcal{R}^k_{top}(u)\right|
\end{equation}
where $k$ is the number of neighbors, $I$ is the set of items, $U$ is the set of users, and $\mathcal{R}^k_{top}(u)$ is the set of top items for user $u$. 
This way, we can test whether parts of the item catalog always receive low estimated rating scores.
We use $\text{PP-Corr}@k$ and $\mathrm{Coverage}@k$ to evaluate \emph{ReuseKNN}$_{DP}$.
Additionally, we also use these metrics to evaluate \emph{UserKNN} to explore the impact of DP~\cite{ekstrand2018privacy}.

\subsection{Datasets}
\label{subsec:datasets}
\begin{table}[!t]
    \centering
    \caption{Descriptive statistics of the five datasets. $|U|$ is the number of users, $|I|$ is the number of items, $|R|$ is the number of ratings, $|R| / |U|$ is the ratings-to-users ratio, $|U| / |I|$ is the users-to-items ratio, and Density is given by $|R|/(|U|\cdot|I|)$.}
    \begin{footnotesize}
    \begin{tabular}{l l r | r r r r r r}
        \toprule
        Dataset & Domain & Rating range & $|U|$ & $|I|$ & $|R|$ & $|R| / |U|$ & $|U| / |I|$ & Density  \\ \midrule
        ML 1M & Movies & \{1\dots5\} & 6,040 & 3,706 & 1,000,209 & 165.60 & 1.6298 & 4.47\%  \\
        Douban & Movies & \{1\dots5\} & 2,509 & 39,576 & 893,575 & 356.15 & 0.0634 & 0.90\%  \\
        LastFM & Music & \{1\dots1,000\} & 3,000 & 352,805 & 1,755,361 & 585.12 & 0.0085 & 0.17\%  \\
        Ciao & Movies & \{1\dots5\} & 7,375 & 105,096 & 282,619 & 38.32 & 0.0702 & 0.04\%  \\ 
        Goodreads & Books & \{1\dots5\} & 20,000 & 508,696 & 2,569,177 & 128.46 & 0.0394 & 0.03\%  \\ \bottomrule
    \end{tabular}
    \end{footnotesize}
    \label{tab:datasets}
\end{table}

In this work, we conduct experiments on five different datasets: \emph{MovieLens 1M} (\emph{ML 1M})~\cite{harper2015movielens}, \emph{Douban}~\cite{hu2014your}, \emph{LastFM User Groups} (\emph{LastFM})~\cite{kowald2020unfairness}, \emph{Ciao}~\cite{guo2014etaf}, and \emph{Goodreads}~\cite{DBLP:conf/recsys/WanM18,DBLP:conf/acl/WanMNM19}.

All five datasets exhibit different properties, as illustrated in Table~\ref{tab:datasets}.
For example, the movie rating dataset \emph{ML 1M} (integer ratings in $\{1\dots5\}$) is the densest dataset.
Similarly, also \emph{Douban} (integer ratings in $\{1\dots5\}$) and \emph{Ciao} (integer ratings in $ \{1\dots5\}$) are movie rating datasets. 
Moreover, in \emph{Ciao}, users have the smallest number of ratings per user (i.e., $|R| / |U|$) on average. 
\emph{LastFM} includes implicit feedback data (i.e., listening counts) from the online music streaming service Last.fm.
However, in this dataset, Kowald et al.~\cite{kowald2020unfairness} transfer the implicit feedback to decimal ratings in $\{1\dots1,000\}$.
Plus, users have the largest number of ratings per users. 
The book rating dataset \emph{Goodreads} (integer ratings in $\{1\dots5\}$), where we use a random sample of 20,000 users, is the largest and least dense dataset. 

Overall, the datasets cover (i) the movie, music, and book domain, (ii) implicit and explicit feedback, and (iii) different descriptive statistics.


\subsection{Evaluation Protocol and Statistical Tests}
\label{subsec:evaluation_protocol}
We perform all experiments using $5$-fold cross-validation, and randomly split all folds into 80\% training sets $R^{train}$ and 20\% test sets $R^{test}$. 
The ratings in $R^{train}$ are used to train the recommendation algorithms, and the ratings in $R^{test}$ represent the rating queries used for evaluation.
Also, we test the statistical significance of our results.
Specifically, after close inspection of our results, we resort to the Mann-Whitney-U-Test.
For the query-based metrics $\mathrm{Neighbors}@q$ and $\mathrm{CoRatings}@q$, we evaluate significance for all rating queries $q \in [2;100]$ when utilizing $k=10$ neighbors.
For other metrics, i.e., $\mathrm{MAE}@k$, $\mathrm{PrivacyRisk}@k$, $\text{PP-Corr}@k$, and $\mathrm{Coverage}@k$, we evaluate significance after all queries have been processed by the recommender system.
Again, here, we utilize $k=10$ neighbors to generate recommendations.
Importantly, throughout this work, we only report statistical significance if we observe significance for each of the five folds.

\subsection{Parameter Settings}
\label{s:parameters}
The proposed approach relies on two adjustable hyperparameters, i.e., (i) the number of neighbors $k$ used in the recommendation process and (ii) the data usage threshold $\tau$. 
To test the performance of \emph{ReuseKNN} and \emph{ReuseKNN}$_{DP}$ for different values of $k$, we vary $k \in \{5, 10, 15, 20, 25, 30\}$.
Plus, we set $\tau$ to the approximate starting value of the tail of \emph{UserKNN}'s data usage distribution $\mathrm{DataUsage@k}$, which is given by its maximal second derivative (see Figure~\ref{fig:approach}).
This way, we assume that only the tail's small privacy risk (as a result of the rare data usage) is tolerable and give an example of how $\tau$ can be defined by the recommender system provider.
Also, $\tau$ is the same for all users.
This leads to the following $\tau$ values for $k=10$: 92.89 (ML 1M), 91.54 (Douban), 104.32 (LastFM), 95.79 (Ciao), and 94.90 (Goodreads). 
For the similarity function $sim$, we use cosine similarity.
\section{Results and Discussion} \label{sec:results}
We structure our results into two parts: 
(i) neighborhood reuse only (\emph{ReuseKNN}), and (ii) neighborhood reuse with DP (\emph{ReuseKNN}$_{DP}$).
\subsection{ReuseKNN}
In this section, we present our evaluation results for \emph{ReuseKNN}, i.e., neighborhood reuse only.

\subsubsection{Neighborhood Reuse}
\label{subsec:neighborhood_growth}
\begin{figure*}[!t]
    \centering
    \includegraphics[width=1\linewidth]{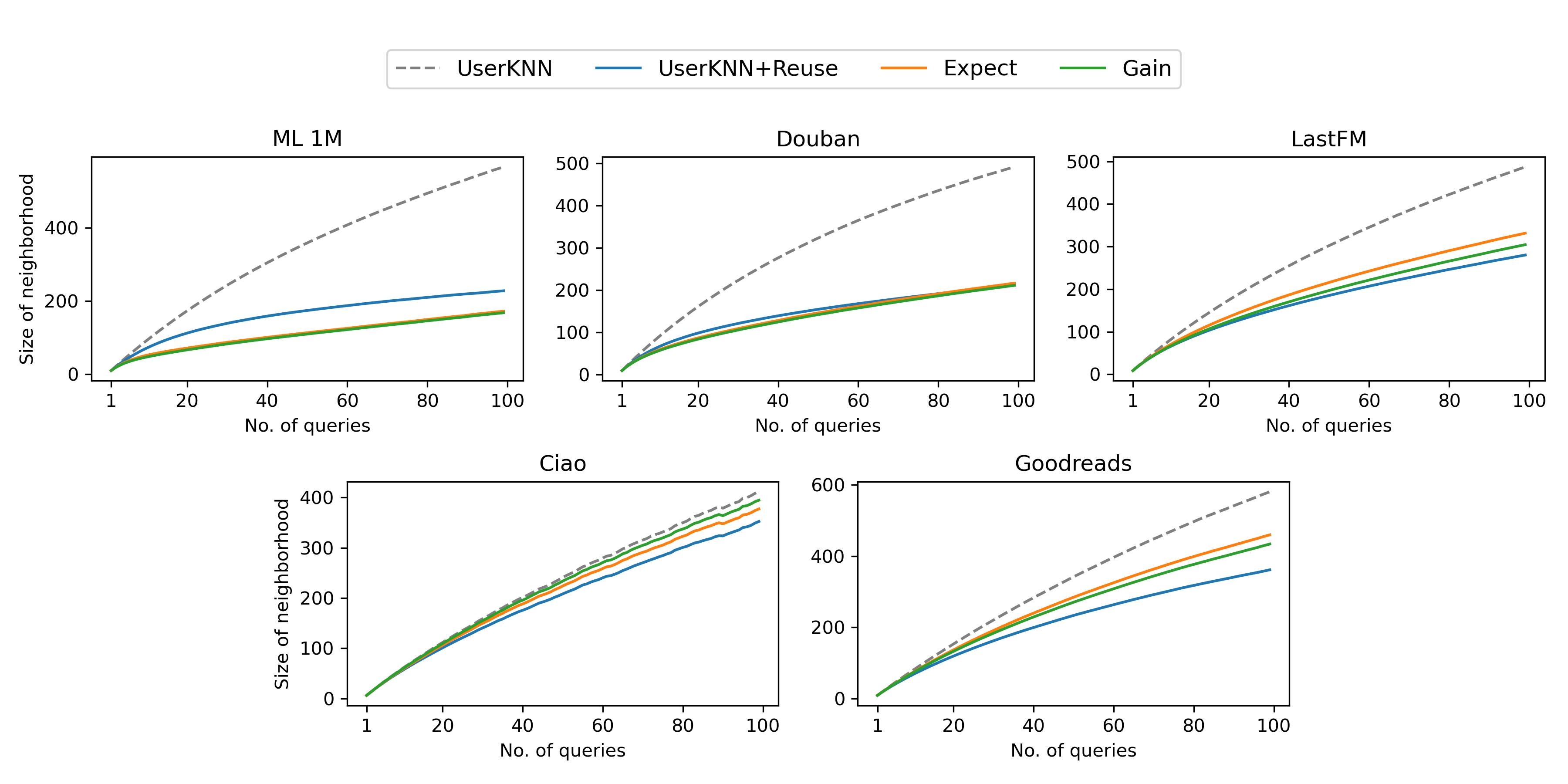}
    \caption{
    Average number of neighbors per target user after $q$ rating queries.
    Our neighborhood reuse strategies utilized in \emph{ReuseKNN}, i.e., \emph{Expect} and \emph{Gain}, generate smaller neighborhoods than \emph{UserKNN}.
    }
    \label{fig:p1_neighborhood_growth}
\end{figure*} 
As the first step in this evaluation stage, i.e., neighborhood reuse only, we investigate the neighborhoods generated by \emph{ReuseKNN}.
Specifically, we compare our neighborhood reuse strategies to our \emph{UserKNN} baseline with respect to the neighborhood size and the number of co-ratings with the target user.
Moreover, we test for statistical significant differences to \emph{UserKNN} after multiple rating queries, i.e., for all $q \in [2; 100]$.

We investigate the average size of target users' neighborhood after $q$ rating queries for a model with $k=10$ neighbors in Figure~\ref{fig:p1_neighborhood_growth}.
For all of our five datasets, 
the size of a user's neighborhood increases more strongly for traditional \emph{UserKNN} than for our neighborhood reuse strategies.
For MovieLens 1M, Douban, LastFM, and Goodreads, a one-tailed Mann-Whitney-U-Test ($\alpha=0.01$) shows that all our neighborhood reuse strategies yield significantly smaller neighborhoods than traditional \emph{UserKNN} for $q \in [2; 100]$ rating queries.
This means that \emph{ReuseKNN} can already reuse neighbors after an initial neighborhood is generated for the very first rating query.

However, for Ciao, multiple initial rating queries are needed to generate reusable neighborhoods.
Our neighborhood reuse strategies tend to yield significantly smaller neighborhoods only for a few rating queries.
For \emph{Gain}, we do not observe significant differences.
We attribute this to the on average small user profiles in Ciao (see Table~\ref{tab:datasets}).
Reusable neighbors are scarce and thus, \emph{ReuseKNN} cannot reduce the neighborhood size as effectively as in the case of the other datasets.


\begin{figure*}[!t]
    \centering
    \includegraphics[width=\linewidth]{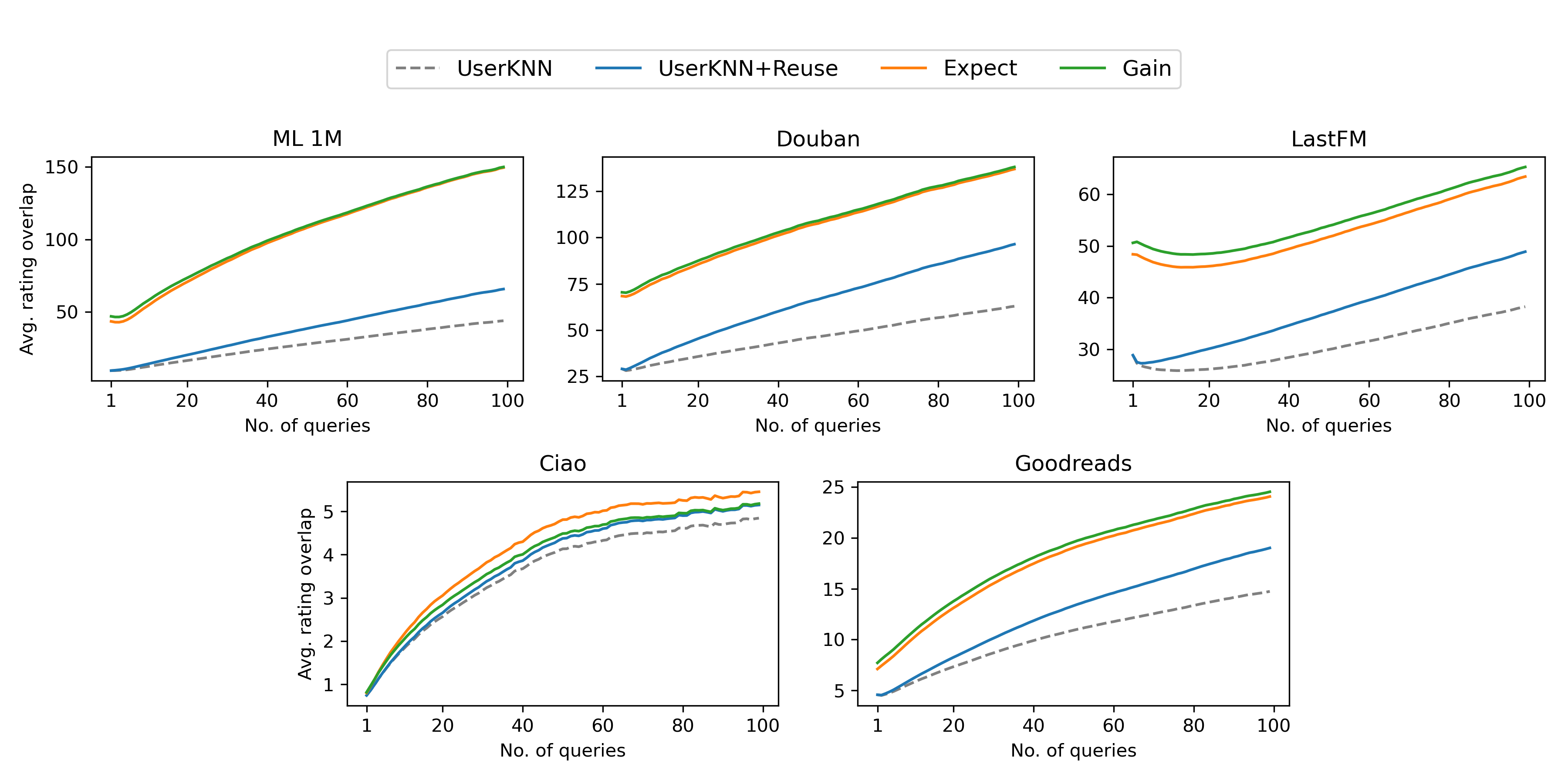}
    \caption{Avg. number of co-rated items between the target user and its neighbors. Our neighborhood reuse strategies for \emph{ReuseKNN}, i.e., \emph{Expect} and \emph{Gain}, generate neighborhoods, in which the neighbors' rated items overlap more with the target users' than in the case of \emph{UserKNN}. 
    With this, neighbors are beneficial for generating accurate recommendations. 
    }
    \label{fig:p1_rating_overlap}
\end{figure*} 

In addition to the neighborhood size, we also investigate the number of co-rated items between the target user and its neighbors after querying $q$ rating queries (see Figure~\ref{fig:p1_rating_overlap}).
Note that as before, the statistical significance is evaluated after multiple rating queries, i.e., for all $q \in [2; 100]$.
For all our five datasets, our neighborhood reuse strategies can substantially increase the number of co-ratings over traditional \emph{UserKNN}.
A one-tailed Mann-Whitney-U-Test ($\alpha=0.01$) reveals that our neighborhood reuse strategies generate neighborhoods with significantly more co-ratings with the target user than \emph{UserKNN} for $q \in [2; 100]$ rating queries.
This indicates that \emph{ReuseKNN} generates neighborhoods with fewer neighbors that have more co-ratings with the target user than in case of traditional \emph{UserKNN}, which can foster recommendation accuracy~\cite{desrosiers2010comprehensive,adomavicius2012impact}.

However, for Ciao, our neighborhood reuse strategies tend to generate neighborhoods with significantly more co-ratings for only a few rating queries.
As in our neighborhood size experiment, we attribute this to the small user profiles in Ciao, which makes neighborhood reuse less effective due to the scarcity of reusable neighbors.


\subsubsection{Accuracy}
\begin{figure*}[!t]
    \centering
    \includegraphics[width=\linewidth]{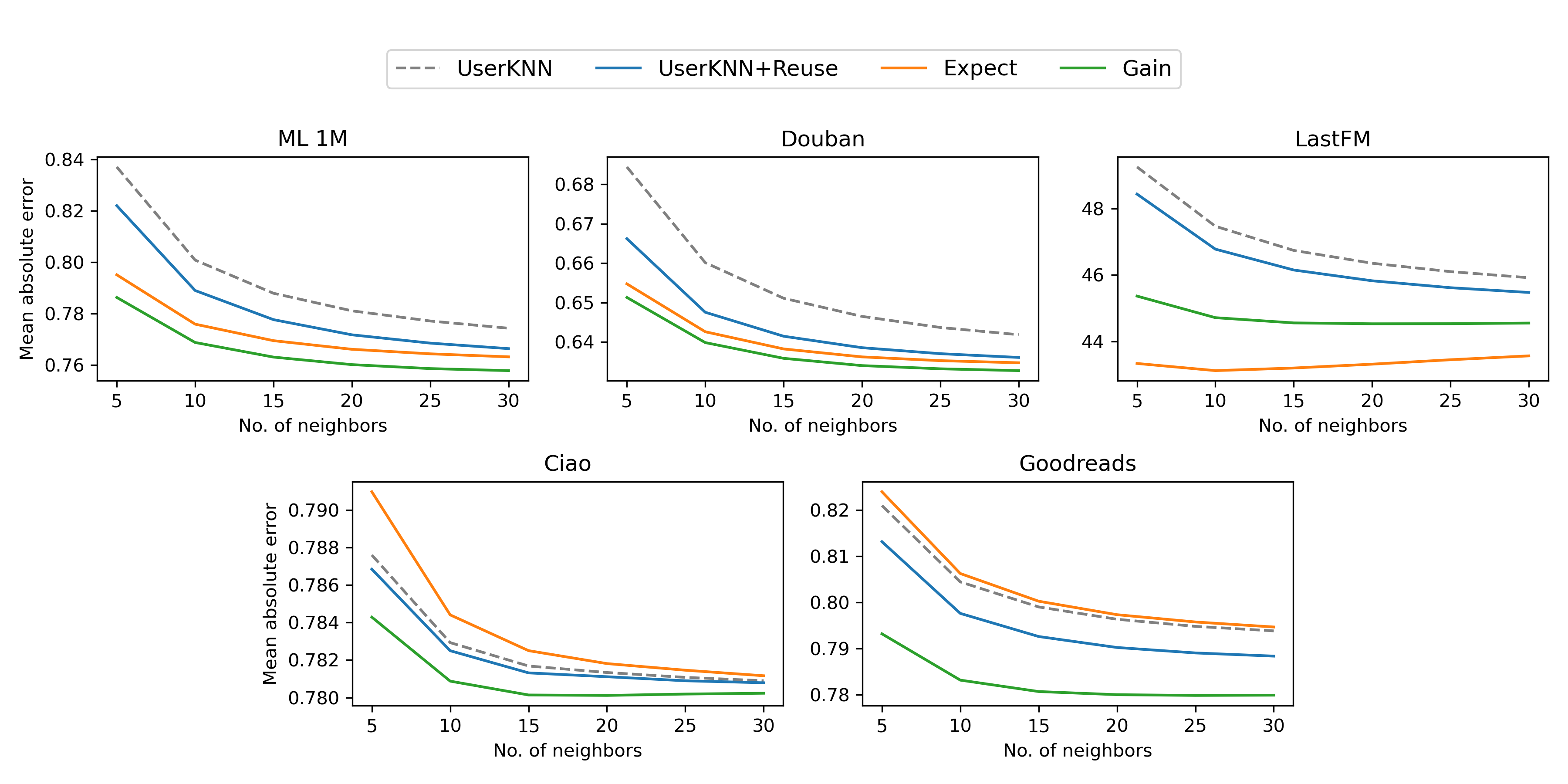}
    \caption{
    Comparison of the recommendation accuracy between \emph{ReuseKNN} and \emph{UserKNN}. \emph{ReuseKNN}'s neighborhood reuse strategies generate more accurate recommendations than \emph{UserKNN}. For sparse datasets (i.e., Ciao and Goodreads), personalized neighborhood reuse (i.e., \emph{Gain}) works better. In contrast, unpersonalized neighborhood reuse (i.e., \emph{Expect}) works better for datasets, in which neighbors are scarce (i.e., LastFM).}
    \label{fig:p1_accuracy}
\end{figure*}

Next, we compare \emph{ReuseKNN} to traditional \emph{UserKNN} in terms of recommendation accuracy (see Figure~\ref{fig:p1_accuracy}).
Specifically, we test for statistical significant differences between our neighborhood reuse strategies and the \emph{UserKNN} baseline.

We find that our neighborhood reuse strategies can generate more accurate recommendations than \emph{UserKNN}. 
This shows that reusing neighbors that have already been used in the past can also lead to meaningful (accurate) recommendations in the future.
Specifically, for ML 1M, Douban, and LastFM, a one-tailed Mann-Whitney-U-Test ($\alpha=0.01$) indicates that our neighborhood reuse strategies significantly increase recommendation accuracy for a model with $k=10$ neighbors. 
Due to personalization, \emph{Gain} performs best across most datasets.

For LastFM, unpersonalized neighborhood reuse (i.e., \emph{Expect}) outperforms personalized neighborhood reuse (i.e., \emph{Gain}).
We attribute this to LastFM's small users-to-items ratio as compared to the other datasets (see Table~\ref{tab:datasets}), which makes it hard to identify neighbors, similar to an item-cold start scenario~\cite{saveski2014item}. 
Concretely, in the case of personalized neighborhood reuse, selecting reusable neighbors for a specific target user reduces the pool of potential neighbors per item to a personalized subset and leads to a worse performance compared to unpersonalized neighborhood reuse.
In contrast, unpersonalized neighborhood reuse allows using the entire pool of potential neighbors and thus achieve a higher accuracy for LastFM. 

In the case of our least dense datasets Ciao and Goodreads, we observe that our personalized neighborhood reuse strategy \emph{Gain} can handle these datasets better than our unpersonalized neighborhood reuse strategy \emph{Expect}. 
\emph{Gain} selects neighbors, whose rating data could have been used by the target user in the past (see Equation~\ref{eq:gain}). 
This way, \emph{Gain} creates a neighborhood for a given target user with sufficient rating data even in sparse datasets.

Plus, we highlight that \emph{Gain} significantly increases recommendation accuracy for Goodreads, despite the dataset's low density. 
In the case of Ciao, a two-tailed Mann-Whitney-U-Test ($\alpha=0.01$) reveals no significant differences between our neighborhood reuse strategies and \emph{UserKNN} for $k=10$, which suggests that all our neighborhood reuse strategies can preserve recommendation accuracy.
As shown in our previous experiments (see Section~\ref{subsec:neighborhood_growth}), neighborhood reuse is less effective for Ciao due to the small user profiles.
Thus, it makes sense that for Ciao, the recommendation accuracy cannot be improved as effectively as for the remaining datasets.


\subsubsection{Percentage of Vulnerable Users}
Previously in Section~\ref{subsec:neighborhood_growth}, we found that neighborhood reuse can significantly reduce the number of neighbors that are utilized in the recommendation process.
Now, however, we analyze how many neighbors are utilized for more than $\tau$ rating queries (i.e., the usage of their data exceeds threshold $\tau$) and thus, need to be protected with DP (see Table~\ref{tab:frac_vulnerables}).
Specifically, we compare our neighborhood reuse strategies to the \emph{UserKNN} baseline.

For all of our five datasets, our neighborhood reuse strategies lead to less vulnerable users than traditional \emph{UserKNN}.
Especially, \emph{Except} shows the best (i.e., lowest) percentage of vulnerable users. 
For example, for the ML 1M dataset, \emph{UserKNN} leads to 80.39\% of users that are vulnerable, since their data usage exceeds threshold $\tau=92.89$ (see Section~\ref{s:parameters}), whereas \emph{Expect} leads to only 24.13\% vulnerable users and thus, fewer users need to be protected with DP.

For Ciao, our neighborhood reuse strategies achieve only minor improvements over \emph{UserKNN}.
The reason is that \emph{UserKNN} already yields a small percentage of vulnerable users and as such, \emph{ReuseKNN} leads to only small improvements.
Additionally, our previous findings show that the effect of neighborhood reuse on Ciao is smaller than on the remaining datasets due to the small average user profile size (see Table~\ref{tab:datasets}). 
This leads to a lack of reusable neighbors and, thus, also limits the effect neighborhood reuse has on the percentage of vulnerable users.

\setlength{\dashlinedash}{3pt} 
\setlength{\dashlinegap}{1.5pt} 

\begin{table}[!t]
    \centering
    \caption{
    Percentage of vulnerable users for a model with $k=10$ neighbors. Best results, i.e., lowest values, are in \textbf{bold}. 
    For all datasets, \emph{ReuseKNN}'s \emph{Expect} neighborhood reuse strategy leads to fewer vulnerable users than \emph{UserKNN}. 
    For Ciao, our neighborhood reuse strategies can achieve only minor improvements, as already \emph{UserKNN} yields a small percentage of vulnerable users.}
    \begin{footnotesize}
    \begin{tabular}{l c c c c c}
        \toprule
        Method & ML 1M & Douban & LastFM & Ciao & Goodreads \\ \midrule
        UserKNN & 80.39\% & 96.68\% & 99.89\% & 8.02\% & 65.00\% \\ 
        UserKNN+Reuse & 84.64\% & 87.37\% & 98.90\% & 7.91\% & 52.29\% \\ \cdashline{1-6}
        Expect & \textbf{24.13\%} & \textbf{34.40\%} & \textbf{68.20\%} & \textbf{7.88\%} & \textbf{29.12\%} \\
        Gain & 25.09\% & 37.43\% & 80.28\% & 8.19\% & 40.51\%  \\
        \bottomrule
    \end{tabular}
    \end{footnotesize}
    \label{tab:frac_vulnerables}
\end{table}

\subsubsection{Summary}
Overall, we find that through neighborhood reuse, \emph{ReuseKNN} can significantly reduce the size of target users' neighborhoods as compared to traditional \emph{UserKNN}. 
Despite the much smaller neighborhoods, \emph{ReuseKNN} identifies neighbors that have many more co-rated items with the target user than in the case of \emph{UserKNN}.
As related work suggests, these neighbors are more ``reliable'' and can be crucial for recommendation accuracy~\cite{desrosiers2010comprehensive,adomavicius2012impact}. 

Based on the much smaller but more reliable neighborhoods, \emph{ReuseKNN} can provide significantly higher recommendation accuracy than traditional \emph{UserKNN}.
For sparse datasets, personalized neighborhood reuse seems to be a better solution than unpersonalized neighborhood reuse.

Plus, \emph{ReuseKNN} can substantially reduce the percentage of vulnerable users, and in general, our \emph{Except} neighborhood reuse method yields the fewest vulnerable users.

\subsection{ReuseKNN$_{DP}$}
Next, we present our results on \emph{ReuseKNN}$_{DP}$, i.e., neighborhood reuse with DP.

\subsubsection{Accuracy}
\begin{figure*}[!t]
    \centering
    \includegraphics[width=\linewidth]{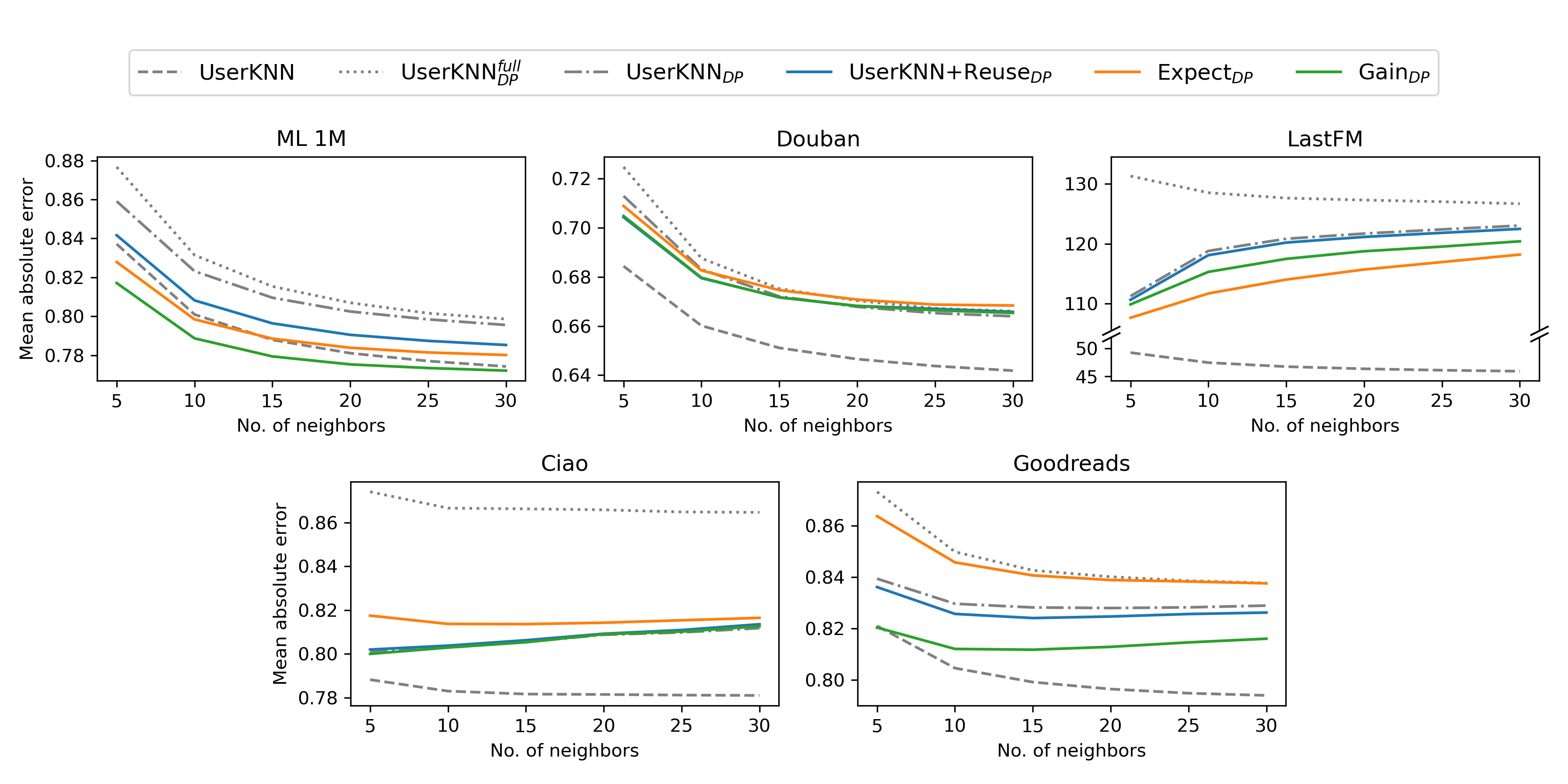}
    \caption{
    Comparison of the recommendation accuracy between \emph{ReuseKNN}$_{DP}$ and \emph{UserKNN}$_{DP}$.
    We find that \emph{ReuseKNN}$_{DP}$'s neighborhood reuse strategies, i.e., \emph{Expect}$_{DP}$ and \emph{Gain}$_{DP}$, can preserve or even improve recommendation accuracy in terms of lower MAE.
    This shows that reducing the number of users to which DP has to be applied can help to increase recommendation accuracy.}
    \label{fig:p2_accuracy}
\end{figure*}

First and foremost, we note that in our experiments without DP (see Figure~\ref{fig:p1_accuracy}), \emph{UserKNN} could be outperformed by \emph{ReuseKNN}.
In our experiments with DP however (see Figure~\ref{fig:p2_accuracy}), it is apparent that all evaluated DP-methods do not reach the accuracy of non-DP \emph{UserKNN}.
This means that in general, due to DP, drops in recommendation accuracy have to be expected.
However, in the next paragraphs, we will investigate whether \emph{ReuseKNN}$_{DP}$ can make this accuracy-drop less severe, compared to using the baselines.
In detail, we compare our neighborhood reuse strategies to the \emph{UserKNN}$_{DP}$ baseline and test for statistically significant differences.
Furthermore, we incorporate \emph{UserKNN} without DP and \emph{UserKNN}$^{full}_{DP}$ as additional baselines for our experiments.

In general, for our neighborhood reuse strategies, DP does not cause an accuracy-drop as severe as in case of \emph{UserKNN}$_{DP}$ (see Figure~\ref{fig:p2_accuracy}).
Plus, as expected, \emph{UserKNN$^{full}_{DP}$} performs worst, due to the randomness that is added via DP to the rating data of all users.
This shows that our neighborhood reuse concept helps to generate accurate recommendations in differentially-private KNN-based recommender systems.
For ML 1M and LastFM, a one-tailed Mann-Whitney-U-Test ($\alpha=0.01$) indicates that our neighborhood reuse strategies significantly increase recommendation accuracy over \emph{UserKNN}$_{DP}$ for a model with $k=10$ neighbors.
Additionally, for ML 1M, \emph{Gain}$_{DP}$ performs better than our non-DP baseline \emph{UserKNN}.

Moreover, we observe that LastFM is highly sensitive to the incorporation of DP, since the mean absolute error magnitudes differ substantially between our non-DP experiment in Figure~\ref{fig:p1_accuracy} and our DP experiment in Figure~\ref{fig:p2_accuracy}.
In line with our previous results on non-DP \emph{ReuseKNN}, also \emph{ReuseKNN}$_{DP}$'s unpersonalized neighborhood reuse strategy \emph{Except}$_{DP}$ cannot increase recommendation accuracy for Ciao and Goodreads, which are our two sparsest datasets. 
However, our personalized neighborhood reuse strategy \emph{Gain}$_{DP}$ generates recommendations with significantly higher accuracy for Goodreads. 
For Ciao, no significant differences are found according to a two-tailed Mann-Whitney-U-Test ($\alpha=0.01$).
Thus, \emph{Gain}$_{DP}$ can preserve recommendation accuracy.

For Douban, we observe no significant differences between our neighborhood reuse strategies and \emph{UserKNN}$_{DP}$.
We found empirically that for Douban, \emph{UserKNN}$_{DP}$ and \emph{ReuseKNN}$_{DP}$ utilize more rating data from vulnerable users, than in case of our remaining datasets.
Thus, we measure the fraction of rating data, each user contributes to the dataset, i.e., $|R_u|/|R|$, where $R$ are all users' ratings and $R_u$ are user $u$'s ratings.
We find that for Douban, the 5\% of users with the largest user profiles contribute substantially more ratings to the dataset than for our other datasets, i.e., 0.0008 (ML 1M), 0.0022 (Douban), 0.0012 (LastFM), 0.0009 (Ciao), and 0.0003 (Goodreads).
This suggests that in the case of Douban, the recommendation process more often utilizes these users due to their abundance of rating data.
This, however, makes these users more vulnerable.
Therefore, we suppose that this strong utilization of DP-protected rating data from vulnerable users leads to no significant differences in accuracy between \emph{UserKNN}$_{DP}$ and \emph{ReuseKNN}$_{DP}$.

For Douban, we additionally compare \emph{ReuseKNN}$_{DP}$ to \emph{UserKNN}$^{full}_{DP}$.
Our results suggest that our personalized reuse strategy \emph{Gain}$_{DP}$ generates recommendations with significantly higher accuracy, while \emph{Except}$_{DP}$ show no significant differences. 
Thus, all our neighborhood reuse strategies can preserve recommendation accuracy for this dataset. 

\subsubsection{Privacy Risk}
\begin{figure*}[!t]
    \centering
    \includegraphics[width=\linewidth]{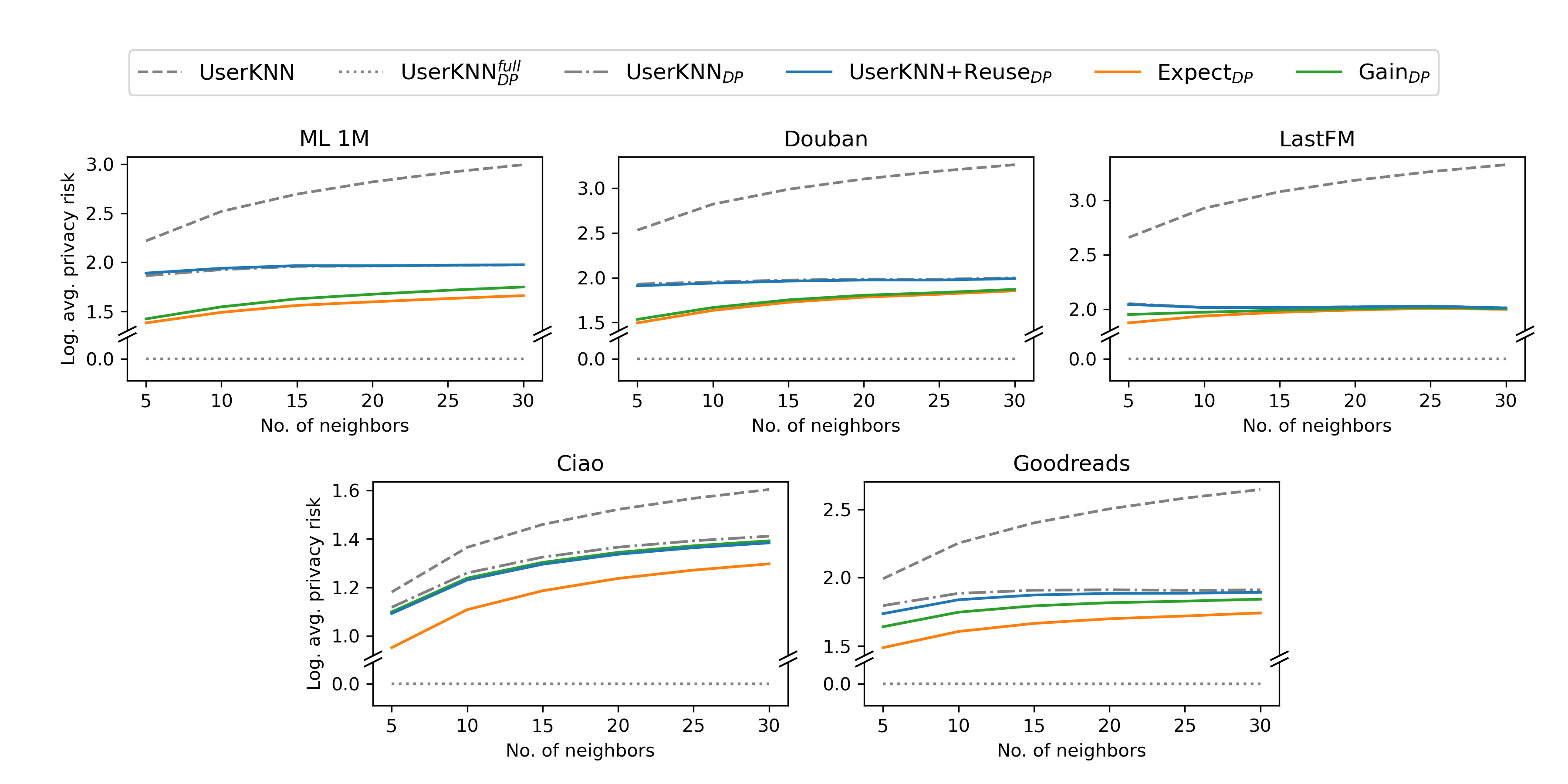}
    \caption{Logarithm (base 10) of the privacy risk averaged over all users. 
    \emph{ReuseKNN}$_{DP}$'s neighborhood reuse strategies yield lower privacy risk than \emph{UserKNN}$_{DP}$.
    This is due to the fact that \emph{ReuseKNN}$_{DP}$ reduces the percentage of users with a privacy risk of $\tau$ (i.e., vulnerables) and simultaneously, decreases the privacy risk of the remaining users (i.e., secures).
    Overall, we find that our unpersonalized neighborhood reuse strategy \emph{Expect}$_{DP}$ achieves the best user privacy, i.e., the lowest privacy risk.} 
    \label{fig:p2_privacy}
\end{figure*}

In \emph{ReuseKNN}$_{DP}$, vulnerable users with high data usage are protected with DP and as such, their privacy risk is set to threshold $\tau$. 
Moreover, secure users' privacy risk is also reduced since they are rarely exploited as neighbors in the recommendation process, i.e., low data usage (see Figure~\ref{fig:approach}).
Specifically, we compare our neighborhood reuse strategies to \emph{UserKNN}$_{DP}$ and test for statistically significant differences.
Furthermore, we use \emph{UserKNN} without DP and \emph{Full}$_{DP}$ as additional baselines.

We visualize the privacy risk of \emph{ReuseKNN}$_{DP}$ and our three baselines \emph{UserKNN}, \emph{UserKNN}$_{DP}$, and \emph{UserKNN}$^{full}_{DP}$ in Figure~\ref{fig:p2_privacy}.
We find that our neighborhood reuse strategies combined with DP can improve user privacy over \emph{UserKNN}$_{DP}$. 
Specifically, a one-tailed Mann-Whitney-U-Test ($\alpha=0.01$) reveals that for our neighborhood reuse strategies on all datasets and for $k=10$, users have significantly less privacy risk than in \emph{UserKNN}$_{DP}$. 

However, for LastFM, this privacy improvement is smaller than for the other datasets. 
Due to the large percentage of vulnerable users for all approaches (see Table~\ref{tab:frac_vulnerables}), most users' privacy risk is set to $\tau$ due to the application of DP.  
Thus, the small percentage of secure users is insufficient to reduce the average privacy risk via neighborhood reuse in the case of LastFM.  

Across all datasets, we observe that our unpersonalized neighborhood reuse strategy \emph{Expect}$_{DP}$ yields the best (lowest) privacy risk. 
This finding is in line with our previous results in Table~\ref{tab:frac_vulnerables}, which show that \emph{Expect}$_{DP}$ performs best with respect to minimizing the percentage of vulnerable users. 
Thus, only a few users have a privacy risk of $\tau$, and the high number of secure users enables to drastically reduce the average privacy risk.
For example, the average privacy risk of secure users for a model with $k=10$ neighbors for \emph{Expect}$_{DP}$ is 11.45 for ML 1M, 18.34 for Douban, 49.92 for LastFM, 15.29 for Ciao, and 18.99 for Goodreads compared to the privacy risk of secure users for \emph{UserKNN}$_{DP}$ that is 50.83 for ML 1M, 62.13 for Douban, 73.42 for LastFM, 21.76 for Ciao, and 41.13 for Goodreads. 
Additionally, a one-tailed Mann-Whitney-U-Test ($\alpha=0.01$) reveals that for ML 1M, Douban, Ciao, and Goodreads, these differences are significant.
Thus, for secure users, \emph{Expect}$_{DP}$ yields a substantially smaller privacy risk than \emph{UserKNN}$_{DP}$.


\subsubsection{Item Popularity Bias}
\begin{table}[!t]
    \centering
    \caption{PP-Corr and Item Coverage for a model with $k=10$ neighbors. 
    Best results, i.e., highest for PP-Corr and lowest for Coverage, are in \textbf{bold}. 
    For PP-Corr, a z-Test~\cite{hinkle2003applied} shows, with ** ($\alpha=0.01$) that our neighborhood reuse strategies as utilized in \emph{ReuseKNN}$_{DP}$ lead to estimated rating scores that are significantly less correlated with item popularity than in case of \emph{UserKNN}$_{DP}$.
    With respect to item coverage, especially \emph{Expect}$_{DP}$ can cover a larger percentage of the item catalog than \emph{UserKNN}$_{DP}$.
    Overall, our results suggest that \emph{ReuseKNN}$_{DP}$ does not increase item popularity bias over \emph{UserKNN}$_{DP}$.
    }
    \begin{adjustbox}{width=\linewidth}
    \begin{tabular}{l r r r r r r r r r r}
    \toprule
    & \multicolumn{2}{c}{ML 1M} & \multicolumn{2}{c}{Douban} & \multicolumn{2}{c}{LastFM} & \multicolumn{2}{c}{Ciao} & \multicolumn{2}{c}{Goodreads} \\ \cmidrule(lr){2-3} \cmidrule(lr){4-5} \cmidrule(lr){6-7} \cmidrule(lr){8-9} \cmidrule(lr){10-11}
    & PP-Corr & Coverage & PP-Corr & Coverage & PP-Corr & Coverage & PP-Corr & Coverage & PP-Corr & Coverage \\ \midrule
    UserKNN & \textbf{0.8405} & 87.94\% & \textbf{0.6780} & 23.50\% & \textbf{0.7339} & 6.11\% & \textbf{0.9755} & 63.19\% & 0.9318 & 29.56\%  \\ \cdashline{1-11}
    UserKNN$_{DP}$ & 0.8742 & 88.77\% & 0.7589 & 26.55\% & 0.8625 & 15.54\% & 0.9758 & 64.03\% & 0.9409 & 31.59\%  \\
    UserKNN$^{full}_{DP}$ & 0.8800 & \textbf{89.53\%} & 0.7675 & 27.65\% & 0.8597 & \textbf{15.86\%} & 0.9778 & \textbf{66.72\%} & 0.9523 & \textbf{34.13\%}  \\ 
    UserKNN+Reuse$_{DP}$ & 0.8750 & 88.37\% & 0.7523 & 27.67\% & 0.8779 & 15.46\% & 0.9759 & 64.26\% & 0.9407 & 31.74\% \\ \cdashline{1-11}
    Expect$_{DP}$ & 0.8688 & 88.83\% & $^{**}$0.7400 & \textbf{28.75\%} & 0.8773 & 14.32\% & 0.9767 & 64.58\% & $^{**}$\textbf{0.9317} & 34.69\% \\
    Gain$_{DP}$ & 0.8725 & 88.07\% & $^{**}$0.7428 & 28.61\% & 0.8621 & 14.77\% & 0.9769 & 64.01\% & 0.9454 & 31.46\% \\
    \bottomrule
    \end{tabular}
    \end{adjustbox}
    \label{tab:popularity_bias}
\end{table}

We test for item popularity bias in \emph{ReuseKNN}$_{DP}$'s recommendations via comparing \emph{ReuseKNN}$_{DP}$ to our \emph{UserKNN}$_{DP}$ baseline with respect to two metrics: Positivity-Popularity Correlation (PP-Corr) and Item Coverage (Coverage).
Plus, we use \emph{UserKNN} without DP and \emph{UserKNN}$^{full}_{DP}$ as additional baselines.
Moreover, in case of PP-Corr, we test for statistically significant differences between our neighborhood reuse strategies and \emph{UserKNN}$_{DP}$. 
First and foremost, for ML 1M, Douban, LastFM, and Ciao, the non-DP baseline \emph{UserKNN} yields lower PP-Corr-values than all remaining methods that use DP.
Similarly, applying DP to only vulnerable users yields lower PP-Corr-values than applying DP to all users in case of ML 1M, Douban, Ciao, and Goodreads.
This fits well to related research~\cite{ekstrand2018privacy} that argues that popularity bias can arise due to the recommender system's inability to personalize recommendations when DP is applied.

However, \emph{ReuseKNN}$_{DP}$ can make the impact of DP on popularity bias less severe, since our neighborhood reuse strategies yield a lower PP-Corr than the DP-baseline \emph{UserKNN}$_{DP}$.
Only for Ciao, no notable differences can be observed.
We investigate this in more detail and find that the neighbors identified by \emph{ReuseKNN}$_{DP}$ rated more distinct items than the neighbors identified by \emph{UserKNN}$_{DP}$. 
As shown by related work on item popularity bias in recommender systems (e.g.,~\cite{kowald2020unfairness,abdollahpouri2019unfairness}), users with a larger user profile size tend to consume less popular items, which leads to less popularity bias.
Due to the small number of ratings per user in Ciao (see Table~\ref{tab:datasets}), which is similar to a user cold-start setting~\cite{lacic2015tackling}, no noteworthy effects on popularity bias can be observed.

In addition to PP-Corr, we also evaluate Coverage, i.e., the percentage of items from the entire item catalog that occur within users' sets of top items.
In general, \emph{UserKNN}$^{full}_{DP}$ tends to give the highest item coverage and non-DP \emph{UserKNN} yields the lowest item coverage.
This makes sense, since \emph{UserKNN}$^{full}_{DP}$ protects all rating data with DP and thus, the estimated rating scores are more random than for the remaining approaches.
This leads to more randomized recommendations, and therefore, to high item coverage~\cite{freyne2013evaluating}.
These randomized recommendations also lead to the fact that in Table~\ref{tab:popularity_bias}, \emph{ReuseKNN}$_{DP}$ cannot reach the item coverage of \emph{UserKNN}$^{full}_{DP}$.
However, more randomized recommendations lead to poorer accuracy than our previous results in Figure~\ref{fig:p2_accuracy} show.

Only in case of LastFM, our neighborhood reuse strategies cover fewer items than \emph{UserKNN}$_{DP}$. 
We underline that these item coverage values are negatively correlated with our accuracy results in Figure~\ref{fig:p2_accuracy}.
This indicates that for LastFM, there is a trade-off between recommendation accuracy and item coverage, similar to the well-known trade-off between precision and recall~\cite{buckland1994relationship}.

\subsubsection{Summary}
Overall, our results are in line with the previously presented results for our non-DP \emph{ReuseKNN}.
Through neighborhood reuse, and thus, reducing the number of users that need to be protected with DP, recommendation accuracy can be preserved, and in many cases even significantly improved over \emph{UserKNN}$_{DP}$. 

Also, our neighborhood reuse strategies used in \emph{ReuseKNN}$_{DP}$ lead to significantly smaller privacy risk than \emph{UserKNN}$_{DP}$.
In particular, unpersonalized neighborhood reuse (i.e., \emph{Except}$_{DP}$) performs best in increasing user privacy.
This shows that the combination of neighborhood reuse and DP provides higher privacy than \emph{UserKNN}$_{DP}$. 

Besides, we find that for \emph{ReuseKNN}$_{DP}$, high estimated rating scores are weaker correlated to item popularity than in case of \emph{UserKNN}$_{DP}$ and that \emph{ReuseKNN}$_{DP}$ can estimate high rating scores for more items than \emph{UserKNN}$_{DP}$.
Thus, \emph{ReuseKNN}$_{DP}$ does not increase item popularity bias.

\subsection{Discussion} \label{subsec:discussion}
\begin{table}[!t]
    \centering
    \caption{
    Mean absolute error and average privacy risk values for our neighborhood reuse strategies that are used in \emph{ReuseKNN}$_{DP}$, i.e., \emph{Expect}$_{DP}$ and \emph{Gain}$_{DP}$ (last two rows), and for the \emph{UserKNN}$_{DP}$ baseline ($k=10$). 
    Also, we perform a one-tailed Mann-Whitney-U-Test ($\alpha=0.01$) and mark (with $^{**}$) significantly better (i.e., lower) values than \emph{UserKNN}$_{DP}$.
    Overall, personalized neighborhood reuse (i.e., \emph{Gain}$_{DP}$) yields the best accuracy and unpersonalized neighborhood reuse (i.e., \emph{Expect}$_{DP}$) gives the lowest privacy risk. 
    For Douban and LastFM, \emph{Expect}$_{DP}$ is well-suited as it yields the highest accuracy and lowest privacy risk. 
    For the remaining datasets, all neighborhood reuse strategies provide a less serious accuracy-privacy trade-off than \emph{UserKNN}$_{DP}$.
    }
    \begin{adjustbox}{width=\linewidth}
    \begin{tabular}{l r r r r r r r r r r}
    \toprule
    & \multicolumn{2}{c}{ML 1M} & \multicolumn{2}{c}{Douban} & \multicolumn{2}{c}{LastFM} & \multicolumn{2}{c}{Ciao} & \multicolumn{2}{c}{Goodreads} \\ \cmidrule(lr){2-3} \cmidrule(lr){4-5} \cmidrule(lr){6-7} \cmidrule(lr){8-9} \cmidrule(lr){10-11}
    & MAE & Privacy R. & MAE & Privacy R. & MAE & Privacy R. & MAE & Privacy R. & MAE & Privacy R. \\ \midrule
    UserKNN & 0.80 & 330.77 & 0.66 & 665.17 & 47.46 & 844.94 & 0.78 & 35.21 & 0.80 & 182.26 \\ \cdashline{1-11}
    UserKNN$_{DP}$ & 0.82 & 84.39 & 0.68 & 89.86 & 118.80 & 103.77 & 0.81 & 27.61 & 0.83 & 75.71 \\
    UserKNN$^{full}_{DP}$ & 0.83 & 0.00  & 0.69  & 0.00  & 128.41  & 0.00 & 0.87 & 0.00  & 0.85 & 0.00  \\ 
    UserKNN+Reuse$_{DP}$ & 0.81  & 87.16 & 0.68 & 87.16  & 118.13  & 103.56  & 0.81 & 26.54  & 0.83 & 68.35  \\ 
    \cdashline{1-11}
    Expect$_{DP}$ & $^{**}$0.80 &  $^{**}$31.03 & 0.68 &  $^{**}$43.25 & $^{**}$111.78 &  $^{**}$86.81 & 0.82 &  $^{**}$21.53 & 0.85 &  $^{**}$40.95 \\
    Gain$_{DP}$ & $^{**}$0.79 &  $^{**}$35.30 & 0.68 &  $^{**}$46.57 &  $^{**}$115.31 &  $^{**}$93.95 & 0.81 &  $^{**}$26.74 &  $^{**}$0.81 &  $^{**}$55.90 \\
    \bottomrule
    \end{tabular}
    \end{adjustbox}
    \label{tab:discussion}
\end{table}


We provide a condensed summary of experimental results (see Table~\ref{tab:discussion}) for all evaluated approaches and all five datasets. 
Specifically, we present the accuracy (i.e., $\mathrm{MAE}@k$) and average privacy risk (i.e., $\mathrm{PrivacyRisk}@k$) values for a model with $k=10$ neighbors. 

Overall, non-DP \emph{UserKNN} results in low MAE, but high privacy risk values.
This shows that approaches without DP sacrifice a user's privacy for recommendation accuracy.
However, our neighborhood reuse strategies with DP provide a less serious trade-off between recommendation accuracy and privacy.
Thus, in the following, we briefly discuss advantages and disadvantages of our neighborhood reuse strategies for all five datasets.

Across our neighborhood reuse strategies that are utilized in \emph{ReuseKNN}$_{DP}$, in general, personalized neighborhood reuse (i.e., \emph{Gain}$_{DP}$) provides the best recommendation accuracy.
Plus, unpersonalized neighborhood reuse (i.e., \emph{Expect}$_{DP}$) yields the lowest privacy risk.
For Douban and LastFM, \emph{Expect}$_{DP}$ performs best in both, accuracy and privacy risk. 
Thus, in this case, \emph{Expect}$_{DP}$ is well suited to provide accurate and private recommendations.
For ML 1M, Ciao, and Goodreads, no neighborhood reuse strategy provides the best result in both evaluation criteria. 
Thus, it depends on the recommender system service provider to decide what strategy could be utilized.

\subsection{Additional Considerations and Experiments}
While our experiments reported so far considered a rating prediction task {as motivated by our problem statement in Section~\ref{sec:problem}} (and accordingly, we measured accuracy using the mean absolute error{~\cite{said2014comparative}), we perform additional experiments with regards to a ranking-based recommendation scenario and a neural-based recommender system.} 
{Due to space limitations, the results of these are detailed in the appendices of this manuscript.}
First, we model a ranking-based recommendation scenario, which is very common today. 
Accordingly, we perform experiments using a ranking-based evaluation metric, i.e., nDCG~\cite{jarvelin2002cumulated}, and report results in Appendix~\ref{sec:a_topn}.
Given the widespread adoption of deep learning techniques in latest recommender systems, we also incorporate neighborhood reuse into a popular neural-based approach, i.e., neural collaborative filtering (NeuCF)~\cite{he2017neural}. 
The approach and results are detailed in Appendix~\ref{sec:a_embeddings}.

Overall, our additional experiments reveal the same pattern of results as discussed above. 
That is, the combination of neighborhood reuse and DP can provide a better trade-off between accuracy and privacy than recommendation methods without neighborhood reuse.
This shows the generalizability of the neighborhood reuse principle for other evaluation scenarios and recommendation algorithms.

\section{Conclusion}
In this work, we investigate the efficacy of neighborhood reuse for differentially-private KNN-based recommendations.
We discuss the proposed approach in a two-stage evaluation procedure: (i) neighborhood reuse only, i.e., \emph{ReuseKNN}, to distill the impact of neighborhood reuse on recommendation accuracy and on the percentage of users that need to be protected with differential privacy, and (ii) neighborhood reuse with differential privacy, i.e., \emph{ReuseKNN}$_{DP}$, to investigate the practical benefit of neighborhood reuse for differentially-private KNN-based recommendations. 
We find that \emph{ReuseKNN} and \emph{ReuseKNN}$_{DP}$ can substantially reduce the number of users that need to be protected with DP, while outperforming related approaches in terms of accuracy.
Also, we highlight that \emph{ReuseKNN}$_{DP}$ effectively mitigates users' privacy risk, as most users are rarely exploited in the recommendation process.
Our work illustrates how to address privacy risks in recommender systems through neighborhood reuse combined with DP.

\vspace{2mm} \noindent \emph{Limitations.} 
We recognize two limitations of the proposed approach:
To quantify the privacy risk, we assume that all pieces of data are equally sensitive.
In reality, disclosing a particular piece of information could pose a different level of privacy risk than disclosing another piece of information~\cite{mehdy2021privacy,knijnenburg2013making}.
{Also, we focus on a neighborhood-based recommender system, specifically user-based \emph{KNN}, instead of neural-based recommender systems. The latter are popular due to their ability to extract and exploit rich user and item representations for generating recommendations. However, traditional algorithms such as user-based \emph{KNN}, have been shown to perform well in a variety of real-world use cases~\cite{dacrema2021}. Plus, neighborhood-based recommender systems have the advantage of providing justifiable recommendations and they incorporate new rating data of users efficiently without requiring a complete retraining of the whole model from scratch~\cite{desrosiers2010comprehensive}. 
Nonetheless, we demonstrate in Appendix~\ref{sec:a_embeddings} that neighborhood reuse can be generalized to neural-based recommender systems, e.g., NeuCF~\cite{he2017neural}.}

\vspace{2mm} \noindent \emph{Future Work.} 
In this work, we evaluated the proposed approach using datasets of three different domains (movies, books, and music), future work will consider additional, more sensitive domains, such as, medicine, finance, insurance, or recruiting.
{We will also incorporate neighborhood reuse into other neural-based recommendation models, e.g., BERT4Rec~\cite{sun2019bert4rec}}.
Plus, we plan to study the impact of the proposed approach, i.e., neighborhood reuse and differential privacy, on individual users' preferences towards long-tail items, e.g., by using the dataset from our previous work on fairness in music recommender systems~\cite{kowald2020unfairness}.
Hence, our long-term plan is to investigate the interaction between privacy and fairness, two key aspects of trustworthy recommender systems. 

\section*{Materials}
The Python-based implementation of our work is publicly available\footnote{\url{https://github.com/pmuellner/ReuseKNN}}.
Also, we provide the source code for generating our sample of the Goodreads dataset. 
All remaining datasets are publicly available as well (see Section~\ref{subsec:datasets}).

\section*{Acknowledgements} 
{This research is funded by the ``DDAI'' COMET Module within the COMET --- Competence Centers for Excellent Technologies Programme, funded by the Austrian Federal Ministry for Transport, Innovation and Technology (bmvit), the Austrian Federal Ministry for Digital and Economic Affairs (bmdw), the Austrian Research Promotion Agency (FFG), the province of Styria (SFG) and partners from industry and academia. The COMET Programme is managed by FFG.
This research received support by the TU Graz Open Access Publishing Fund}, the Austrian Science Fund (FWF): P33526 and DFH-23; and by the State of Upper Austria and the Federal Ministry of Education, Science, and Research, through grant LIT-2020-9-SEE-113. {We thank the anonymous reviewers and the associate editor for their valuable remarks and suggestions.}

%

\bibliographystyle{ACM-Reference-Format}
\bibliography{bibliography}

\newpage
\appendix
\section{Detailed Differential Privacy Analysis}
\label{sec:a_privacyanalysis}
Our differential privacy analysis relies on the fact that, even if the adversary is able to infer the rating used in the recommendation process, it is unaware whether this rating is the neighbor's real rating or was randomly generated by our $m_{DP}$ mechanism.
Formally:
\begin{align}
    \frac{Pr[\text{Adversary's assumption: Real rating | Truth: Real rating}]}{Pr[\text{Adversary's assumption: Real rating | Truth: Random rating}]} &= \\ 
    \frac{Pr[\text{Non-DP rating}] + Pr[\text{Real rating | DP rating}] \cdot Pr[\text{DP rating}]}{Pr[\text{Random rating | DP rating}] \cdot Pr[\text{DP rating}]} &= \\
    \frac{Pr[\text{Non-DP rating}]}{Pr[\text{Random rating | DP rating}] \cdot Pr[\text{DP rating}]} + \underbrace{\frac{Pr[\text{Real rating | DP rating}]}{Pr[\text{Random rating | DP rating}]}}_{\text{$m_{DP}$ mechanism}} &= \\
    \frac{1}{0.25} \cdot \frac{Pr[\text{Non-DP rating}]}{Pr[\text{DP rating}]} + \frac{0.75}{0.25} &= \\
    4 \cdot \frac{\frac{\mathrm{PrivacyRisk@k(u)}}{\mathrm{DataUsage@k(u)}}}{\frac{\mathrm{DataUsage@k(u)} - \mathrm{PrivacyRisk@k(u)}}{\mathrm{DataUsage@k(u)}}} + 3 &= \\
    4 \cdot \frac{\mathrm{PrivacyRisk}@k(u)}{\mathrm{DataUsage}@k(u)-\mathrm{PrivacyRisk}@k(u)} + 3 &\leq e^\epsilon  \\
\end{align}
which leads to a privacy parameter of 
\begin{equation}
    \epsilon = \ln \Bigg(3 + 4 \cdot \frac{\mathrm{PrivacyRisk}@k(u)}{\mathrm{DataUsage}@k(u)-\mathrm{PrivacyRisk}@k(u)}\Bigg)
    \label{eq:a_epsilon}
\end{equation}
In case of \emph{UserKNN}$^{full}_{DP}$, all ratings of a user $u$ are protected with DP and therefore, $\mathrm{PrivacyRisk}@k(u) = 0$ which leads to $\epsilon = \ln 3$. 
In case of \emph{UserKNN}, no DP is applied at all and thus, computing $\epsilon$ is not possible since $\epsilon$ is part of the DP-framework.
Therefore, we set $\epsilon = \infty$.
In case of \emph{UserKNN}$_{DP}$ and \emph{ReuseKNN}$_{DP}$, DP is applied to the rating data of users, for which the usage of their data exceeds threshold $\tau$.
Assuming that $u$ is vulnerable, then $\mathrm{DataUsage}@k(u) > \tau$ and $\mathrm{PrivacyRisk}@k(u) = \min [\tau, \mathrm{DataUsage}@k(u) ]$.
Therefore, it follows that $0 < \mathrm{PrivacyRisk}@k(u) < \mathrm{DataUsage}@k(u)$. 
Varying $\mathrm{PrivacyRisk}@k(u)$ within these boundaries yields: 
\begin{equation}
    \ln 3 < \ln \Bigg(3 + 4 \cdot \frac{1}{\mathrm{DataUsage}@k(u)-1}\Bigg) \leq \epsilon \leq \ln \Bigg(3 + 4 \cdot (\mathrm{DataUsage}@k(u)-1)\Bigg) < \infty
    \label{eq:a_epsilon_fix_tau}
\end{equation}
This shows that \emph{UserKNN}$_{DP}$ and \emph{ReuseKNN}$_{DP}$ provide better privacy than \emph{UserKNN}, but worse privacy than \emph{UserKNN}$^{full}_{DP}$.

Moreover, via neighborhood reuse, \emph{ReuseKNN}$_{DP}$ utilizes a vulnerable user $u$ more often as neighbor (with DP-protected data) than \emph{UserKNN}$_{DP}$ does.
Also, note that the privacy risk of $u$ is the same for \emph{ReuseKNN}$_{DP}$ and \emph{UserKNN}$_{DP}$.
From these observations and Equation~\ref{eq:a_epsilon}, we see that the $\epsilon$ value for \emph{ReuseKNN}$_{DP}$ is smaller than the $\epsilon$ value for \emph{UserKNN}$_{DP}$.
Thus, for vulnerable users, our neighborhood reuse principle leads to \emph{ReuseKNN}$_{DP}$ providing better privacy than \emph{UserKNN}$_{DP}$.


\section{Evaluation of Top-N Recommendations}
\label{sec:a_topn}
\begin{figure*}[!t]
    \centering
    \includegraphics[width=1\linewidth]{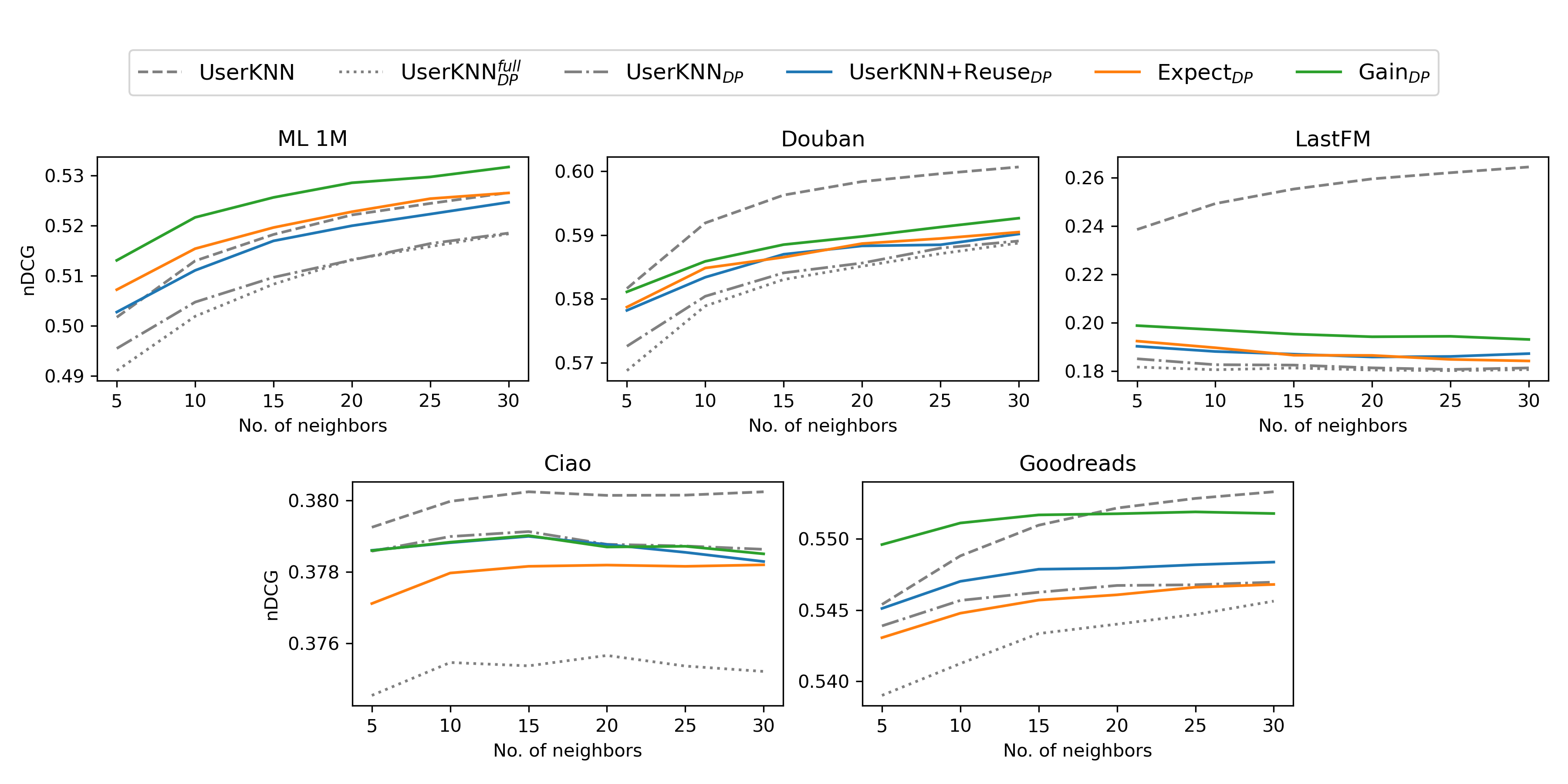}
    \caption{
    nDCG values of each user's top 10 items. The pattern matches our results reported in Section~\ref{sec:results}, i.e., \emph{ReuseKNN}$_{DP}$ can yield better accuracy than \emph{UserKNN}$_{DP}$. Also, especially personalized neighborhood reuse (i.e., \emph{Gain}$_{DP}$) can preserve accuracy well.
    }
    \label{fig:a_ndcg}
\end{figure*} 
In our manuscript, we show that \emph{ReuseKNN}$_{DP}$ can achieve better accuracy in terms of the rating prediction metric MAE than a traditional \emph{KNN} recommender system with DP.
In the following, we evaluate \emph{ReuseKNN}$_{DP}$ in a top-$n$ items recommendation setting via the ranking-aware metric \emph{nDCG} (Normalized Discounted Cumulative Gain)~\cite{jarvelin2002cumulated}.

\subsection{Evaluation Process}
To generate a list of recommended items that can be evaluated via \emph{nDCG}, we select the $n=10$ items with the highest predicted rating score for a given target user $u$~\cite{said2014comparative}.
Formally, for a recommender system model $\mathcal{R}$ and $k$ neighbors, a user $u$'s top-$n$ items are given by:
\begin{equation}
    \mathcal{R}^k_{top}(u) =\ \stackrel{n}{\argmax_{i \in Q_u}} \mathcal{R}^k(u, i)
\end{equation}
where $Q_u$ are the items in $u$'s query set. 
We consider items in the test set as relevant if their true rating exceeds the average rating in the training set of the given dataset.

\subsection{Experiments}
Our results reveal that \emph{Expect}$_{DP}$ and \emph{Gain}$_{DP}$ can yield higher $nDCG$ scores than \emph{UserKNN}$^{full}_{DP}$ (see Figure~\ref{fig:a_ndcg}).
In case of the ML 1M dataset, \emph{Expect}$_{DP}$ and \emph{Gain}$_{DP}$ can even outperform the non-DP baseline \emph{UserKNN}.
Especially \emph{Gain}$_{DP}$ yields high $nDCG$ scores.
Overall, this experiment validates the results of our rating prediction evaluation setting also in a top-$n$ items recommendation setting.


\section{Evaluation of Neural-Based Recommendations}
\label{sec:a_embeddings}
This work considers rating data as input to the recommender system.
However, recommender systems can also use more complex representations of users and items, i.e., embeddings as generated by neural network architectures.
Therefore, in the following, we demonstrate the generalizability of our approach for neural-based recommendation methods. 

\subsection{Generation of Embeddings}
To generate user and item embeddings, we rely on a simple approach inspired by the NeuCF~\cite{he2017neural} architecture.
Specifically, for user $u$ and item $i$, the predicted rating score $y_{u, i}$ is given by: 
\begin{equation}
    y_{u, i} = b + ReLu(w x_u W_u^T x_i W_i)
\end{equation}
where $x_u$ is the id of user $u$, $x_i$ is the id of item $i$, the size of the embedding layer is $d = 16$, $W_{u}, W_{i} \in \mathbbm{R}^d$, $w, b \in \mathbbm{R}$, and $ReLu$ is the activation function.
We apply Adam~\cite{kingmaadam} with a step size of $\alpha=0.001$ to minimize the MAE between $y_{u, i}$ and the rating $r_{u, i}$. 
The parameters $\alpha$ and $d$ are set to the values proposed in~\cite{he2017neural}.
We train the network for 50 epochs and use a batch size of 128. 
Furthermore, we stop training if there is no improvement of the training objective for more than 10 epochs.
After training, the user and item embeddings are given by $x_u W_u$ and $x_i W_i$ respectively.

\subsection{Neural-Based Recommendations}
For our neural-based variants of \emph{UserKNN}, i.e., \emph{NeuKNN} and \emph{NeuKNN}$_{DP}$, we calculate the similarity between the target user and the candidate neighbors based on their user embeddings (see Equation~\ref{eq:userknn_neighborhood}).
For \emph{NeuKNN+Reuse}$_{DP}$, i.e., a embedding-based variant of \emph{ReuseKNN}$_{DP}$, we also use an embedding-based similarity.
Plus, we employ a modified definition of $reusability$ that measures the reusability of a candidate neighbor $c$ based on the previous $t-1$ rating queries of target user $u$:
\begin{equation}
    reusability(c|u, i, t) = \sum_{j \in Q^{(t-1)}_u} \mathbbm{1}_{N_{u, j}}(c) \cdot sim(i, j)
\end{equation}
where $\mathbbm{1}_{N_{u, j}}(c)$ is the indicator function of candidate neighbor $c$ being in $N_{u, j}$.
The item similarity $sim$ is the cosine similarity between $i$'s and $j$'s item embeddings.
Therefore, $reusability(c|u, i, t)$ is the summed-up item similarity between the target item $i$ and all items $j \in Q^{(t-1)}_u$ (i.e., the previous $t-1$ rating queries of $u$) for which $c$ has been used as neighbor.

\subsection{Experiments}
In our experiments, we perform evaluation according to the following procedure:
First, we randomly split the dataset into 5 equally-sized subsets, i.e., $D_{1 \leq i \leq 5}$.
We select $D_1$ and equally partition it into the validation data that is used for validating the user and item embeddings, and the test data that is used for evaluating the recommendations.
The remaining data, i.e., $\bigcup_{2 \leq i \leq 5} D_i$, is used to train the user and item embeddings, and to generate recommendations.
Next, we select $D_i$ and repeat this procedure for all $D_{2 \leq i \leq 5}$. 
Eventually, we compute the mean of our evaluation results.

\begin{figure*}[!t]
    \centering
    \includegraphics[width=1\linewidth]{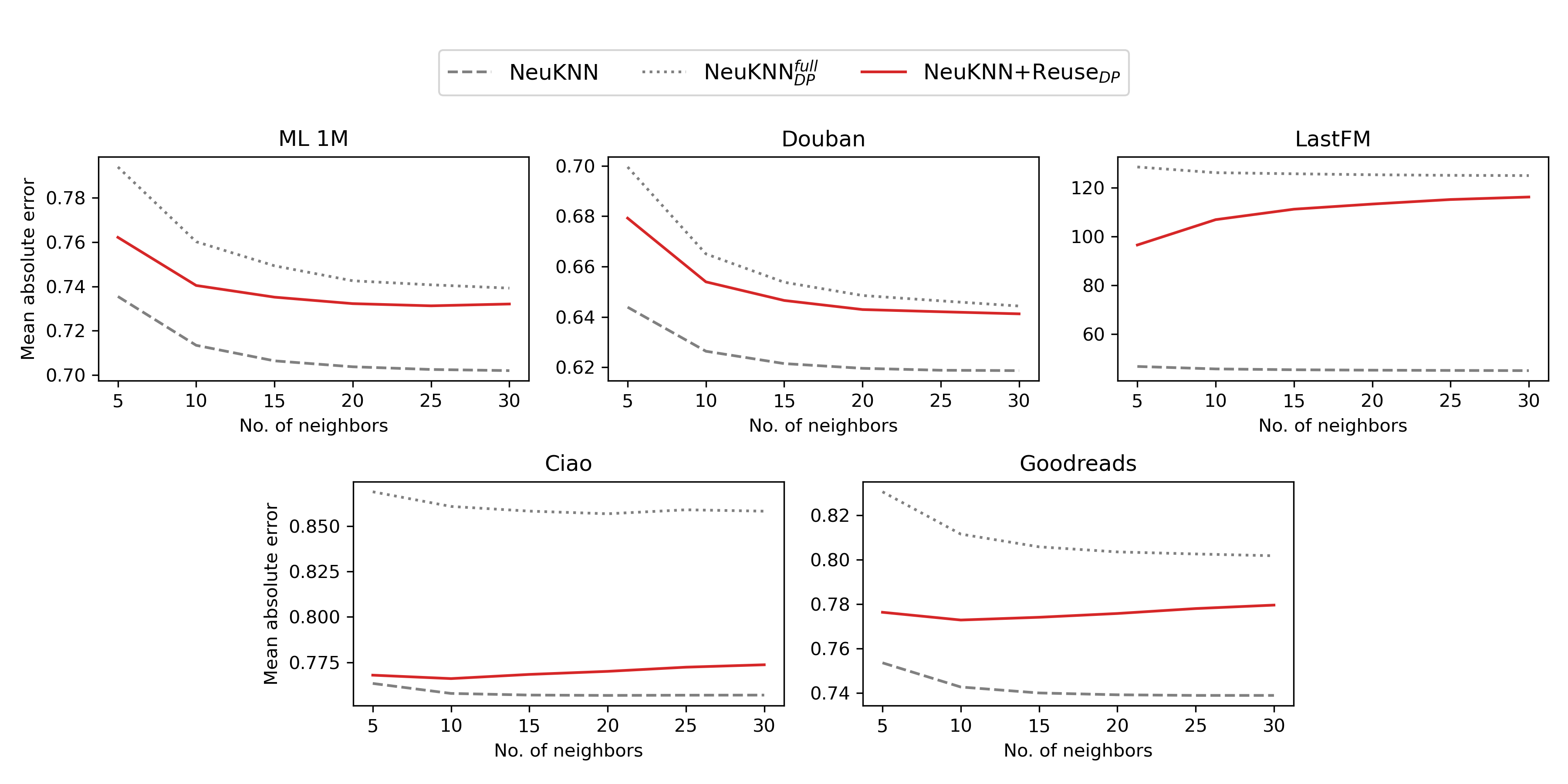}
    \caption{
    Mean absolute error of our neural-based \emph{KNN} recommender system variants.
    Our results indicate that combining neighborhood reuse with DP (i.e., \emph{NeuKNN+Reuse}$_{DP}$) yields better accuracy (lower MAE) than neural-based methods that apply DP without neighborhood reuse (i.e., 
    {\emph{NeuKNN}$^{full}_{DP}$).}
    }
    \label{fig:a_accuracy}
\end{figure*}

\vspace{2mm} \noindent \emph{Accuracy.}
For all datasets, \emph{NeuKNN+Reuse}$_{DP}$ outperforms our baseline \emph{NeuKNN}$^{full}_{DP}$ that applies DP to all users (see Figure~\ref{fig:a_accuracy}).
For completeness, we also visualize \emph{NeuKNN} that does not apply DP at all and thus, yields higher accuracy than both DP-based methods. 
Overall, the result for our embedding-based methods \emph{NeuKNN}$^{full}_{DP}$ and \emph{NeuKNN+Reuse}$_{DP}$ are in line with the results of our rating-based methods, i.e., that the combination of neighborhood reuse and DP yields better accuracy on all five investigated datasets than traditional DP-based methods.

\begin{figure*}[!t]
    \centering
    \includegraphics[width=1\linewidth]{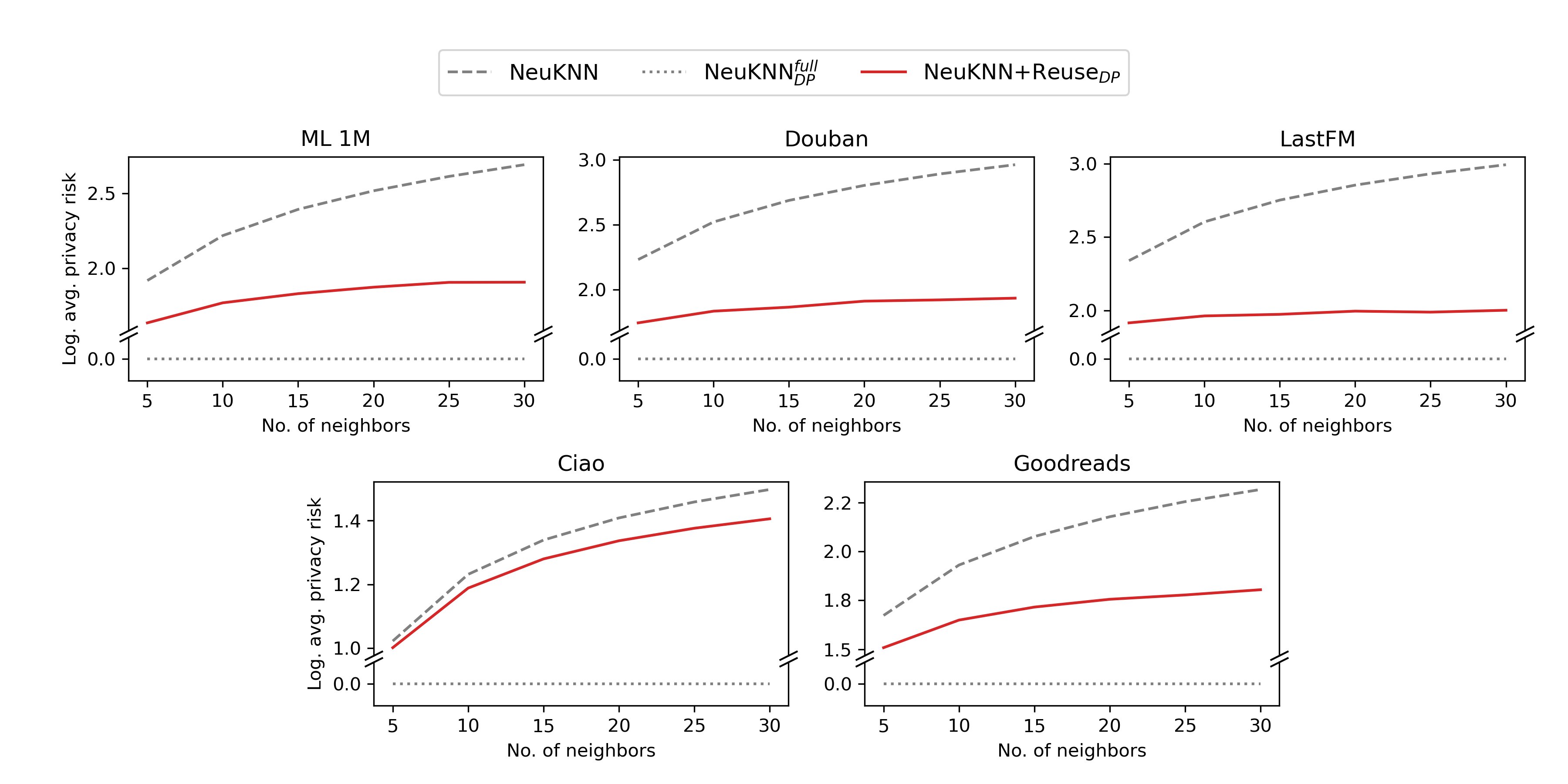}
    \caption{
    Logarithmic (base 10) average privacy risk of our neural-based \emph{KNN} recommender system variants.
    Via combining neighborhood reuse and DP, \emph{NeuKNN+Reuse}$_{DP}$ decreases the users' average privacy risk compared to neural-based methods that do not apply DP (i.e., \emph{NeuKNN}).
    }
    \label{fig:a_privacy}
\end{figure*}

\vspace{2mm} \noindent \emph{Privacy.}
Our baseline \emph{NeuKNN} without DP yields the worst privacy risk, whereas \emph{NeuKNN}$^{full}_{DP}$ yields a privacy risk of zero since all users are protected with DP (see Figure~\ref{fig:a_privacy}).
\emph{NeuKNN+Reuse}$_{DP}$ protects only vulnerable users with DP and in this way, its privacy risk lies between our two baselines.
Therefore, also in terms of privacy risk, the results of our embedding-based experiments match the pattern of the results of our rating-based methods.

\end{document}